\documentclass{aa}  
\usepackage{txfonts}
\usepackage[toc,page]{appendix}
\usepackage[dvipsnames,svgnames,table]{xcolor}
\usepackage[utf8]{inputenc}
\usepackage[T1]{fontenc}
\usepackage{hyperref}
\usepackage{url}
\usepackage{booktabs}
\usepackage{amsfonts}
\usepackage{nicefrac}       
\usepackage{microtype}      
\usepackage{footmisc}
\usepackage{graphicx}
\usepackage{natbib}
\usepackage{tabu}
\usepackage{makecell}
\usepackage{listings}
\usepackage{amsmath}
\usepackage{multirow}

\usepackage{multiobjective}

\usepackage{xspace}
\newcommand{\alphapmp}{$\alpha$PMP\xspace}
\newcommand{\tefflogg}{$T_\mathrm{eff}$ and $\log g$\xspace}

\def\shrug{\texttt{\raisebox{0.75em}{\char`\_}\char`\\\char`\_\kern-0.5ex(\kern-0.25ex\raisebox{0.25ex}{\rotatebox{45}{\raisebox{-.75ex}"\kern-1.5ex\rotatebox{-90})}}\kern-0.5ex)\kern-0.5ex\char`\_/\raisebox{0.75em}{\char`\_}}}

\usepackage{tikz}
\usetikzlibrary{positioning, arrows, shapes, calc}
\usetikzlibrary{shapes.multipart}
\usepackage{upgreek}

\newcommand{\CEMPprecision}{\( 0.96 \pm 0.09 \)\xspace}
\newcommand{\CEMPrecall}{\( 0.68 \pm 0.05 \)\xspace}
\newcommand{\aPMPprecision}{\( 0.84 \pm 0.04 \)\xspace}
\newcommand{\aPMPrecall}{\( 0.11 \pm 0.02 \)\xspace}

\newcommand{\corrFE}{\ensuremath{0.92\pm0.01}\xspace}
\newcommand{\corrC}{\ensuremath{0.92\pm0.01}\xspace}
\newcommand{\corra}{\ensuremath{0.82\pm0.02}\xspace}

\begin{document} 
   \title{Towards model-free stellar chemical abundances}
   \subtitle{Potential applications in the search for chemically peculiar stars in large spectroscopic surveys}
   \author{Theosamuele Signor
          \inst{1,2}\thanks{\email{theosamuele.signor@mail.udp.cl}}
          \and
          Paula Jofré\inst{1}
          \and
          Hernan Lira\inst{2}
          \and
          Sara Vitali\inst{1}
          \and
          Luis Martí\inst{2}
          \and
          Nayat Sánchez-Pi\inst{2}
          }
   \institute{Instituto de Estudios Astrofísicos, Facultad de Ingeniería y Ciencias, Universidad Diego Portales, Av. Ejercito 441, Santiago, Chile
         \and
         Inria Chile Research Center, Av. Apoquindo 2827, piso 12, Las Condes, Santiago, Chile
             }
   \date{Received 2 May 2025; accepted 11 Nov 2025}

  \abstract
  {Chemical abundance determinations from stellar spectra are challenged by observational noise, limitations in stellar models, and departures from simplifying assumptions. While traditional and supervised machine learning methods have made remarkable progress in estimating atmospheric parameters and chemical compositions within existing physical models, these factors still constrain our ability to fully exploit the vast data sets provided by modern spectroscopic surveys.}
   {We aim to develop a self-supervised, disentangled representation learning framework that extracts chemically meaningful features directly from spectra, without relying on externally imposed label catalogs.}
   {We build a variational autoencoder-based representation learning model with physics‐inspired structure: multiple decoders each focus on spectral regions dominated by a particular element, enforcing that each latent dimension maps to a single abundance. 
   To evaluate the potential application of our framework, we trained and validated the model on low-resolution, low signal-to-noise synthetic spectra focusing on $\rm [Fe/H]$, $\rm [C/Fe]$, and $\rm [\alpha/Fe]$. We then demonstrate how the trained model can be used to flag stars as chemically enhanced or depleted in these abundances based on their position within the latent distribution.}
  {Our model successfully learns a representation of spectra whose axes correlate tightly with the target abundances ($r=\corrFE$ for $\rm [Fe/H]$, $r=\corrC$ for $\rm [C/Fe]$, $r=\corra$ for $\rm [\alpha/Fe]$).
  The disentangled representations provide a robust means to distinguish stars based on their chemical properties, offering an efficient and scalable solution for large spectroscopic surveys.}
   {}
   \maketitle
\section{Introduction}\label{sec:intro}
Extracting reliable chemical information from stellar spectra has long been a central challenge in astrophysics, especially given the vast datasets produced by modern spectroscopic surveys. Large-scale programs—including the Gaia-ESO Survey \citep{gilmore2022gaia,randich2022gaia}, the  Galactic Archaeology with HERMES (GALAH) Survey \citep{desilva2015galah,buder2021galah}, the Dark Energy Spectroscopic Instrument (DESI), the Sloan Digital Sky Survey (SDSS; \citealt{Smee2013multi, Kollmeier2019sdss}), in particular the Apache Point Observatory Galactic Evolution Experiment (APOGEE, \citealt{Majewski2017Apache}) and the Baryon Oscillation Spectroscopic Survey (BOSS; \citealt{Dawson2013baryon})—are providing enormous numbers of stellar spectra that enable comprehensive studies of stellar populations and Galactic evolution. However, analyzing these datasets is challenging due to imperfect theoretical models, observational uncertainties, and the difficulty of ensuring consistent abundance determinations across different surveys \citep{jofre2019accuracy}.   

In this era, supervised machine learning methods—such as The Cannon \citep{ness2015cannon}, AstroNN \citep{leung2019deep}, and Payne \citep{Ting2019payne}—have become widely used for chemical analysis, with many chemical abundance catalogs now produced using neural networks trained on synthetic or observed spectral grids.
Yet, these methods still depend heavily on high-quality training data, typically derived from traditional template-fitting techniques. Their performance is sensitive to the quality, coverage, and representativeness of the training sets, and they often struggle with rare or chemically peculiar stars \citep[\emph{e.\,g.},][]{Goodfellow2016deep,Li2025machine}. Also, astrophysical correlations in the training data can lead to biases, where inferred abundances may reflect indirect associations with stellar parameters rather than the true chemical properties \citep[\emph{e.\,g.},][]{ting2024why}. More fundamentally, because the training labels themselves are based on theoretical models rather than direct observational ground truth, these approaches remain model-driven at their core, incorporating assumptions about the underlying physics of stellar atmospheres indirectly through their training data sets.
As a result, neither theoretical models nor, consequently, supervised learning models fully capture the information contained in stellar spectra.

Addressing these challenges requires novel approaches that mitigate biases and enhance generalization across diverse stellar populations. Unsupervised learning techniques offer a promising avenue, as they do not depend on extensive labeled training samples and can uncover hidden structures in spectral data on their own. For example, Autoencoders \citep[AEs,][]{Goodfellow2016deep} are deep neural networks trained to reconstruct their input, a self-supervised strategy \citep[see, \emph{e.\,g.},][]{Gui2023survey} that enables them to compress high-dimensional spectra into compact latent spaces. This reconstruction objective can allow the model to capture intrinsic chemical variations while mitigating the impact of noise \citep[\emph{e.\,g.},][]{Fang2021variational, Carbajal2021disentanglement}.

In this work, we propose a self-supervised deep learning framework for representation learning in stellar spectra designed to learn low-dimensional representations where latent variables exhibit explicit relationships with chemical abundances. 
Consequently, this structured representation is physically interpretable and facilitates downstream tasks such as chemical abundance estimation.

The Milky Way’s stellar halo provides an ideal environment for applying these unsupervised methods. As a relic of past accretion events, the halo preserves the chemical imprints of disrupted dwarf galaxies and clusters \citep[\emph{e.\,g.},][]{2004Venn,2008Delucia,2024Belokurov}. While their spatial coherence is eventually erased, chemical abundances remain intact, offering a unique proxy to reconstruct the accretion history of the Milky Way.  
For example, dwarf galaxies, with their slow star formation and small gravitational potentials, are typically metal-poor and exhibit a lower ratio of $\alpha$-elements to iron \citep[\emph{e.\,g.},][]{2015Hudson,2023Johnson}. In the metal-poor environment of the halo, a significant fraction of stars exhibit enhanced carbon abundances \citep{2013Yong,2014Placco,2018Yoon}. These high carbon abundances at low metallicities suggest enrichment from multiple astrophysical sources, such as massive stars, binary interactions, and various types of supernovae \citep{Beers2005discovery}. Among  these, some show carbon enhancement without signs of s-process enrichment or binarity \citep{2013Norris,2016Hansen,2019Arentsen}. These stars likely inherited their carbon from the early interstellar medium enriched by the first generations of stars \citep{2013Spite,2018Sharma,2019Hartwig}.

Overall, studying the chemical signatures of resolved stars in the halo can be used for tracing the first generations of stars, understanding the role of accreted systems in shaping the Milky Way, and reconstructing the Galaxy’s nucleosynthetic and accretion history \citep{Frebel2007probing,FrebelNorris2015near, Helmi2020}. 
We demonstrate the effectiveness of our approach on low-resolution synthetic stellar spectra, exploring its potential for identifying chemically enhanced or depleted stars—specifically, $\alpha$-poor, metal-poor (\alphapmp) and carbon-enhanced, metal-poor (CEMP) stars—without requiring explicit labels. 
By applying deep representation learning in a self-supervised setting, our approach offers a scalable, data-driven tool for exploring the chemical diversity of the Milky Way.

\subsection{Related work}
A variety of approaches—including template-based, hybrid, and machine learning methods—have been developed to extract chemical information from low-resolution spectra. For example, \cite{lee2011segue,lee2013carbon} developed techniques to determine carbon abundances in metal-poor stars using SDSS spectra, while \cite{Xiang2019abundance} estimated $\alpha$-element abundances from LAMOST spectra. Similarly, \cite{Marsteller2009automated} employed automated spectral synthesis to infer metallicities and carbon abundances, and \cite{bu2016elm} used an Extreme Learning Machine \citep{huang2006extreme} to estimate $\alpha$-element abundances. Although these methods demonstrate the potential of low-resolution spectroscopic data for chemical abundance analysis, they rely on pre-determined templates or labeled training sets, which may limit their applicability to rare or unusual spectral types.

Beyond traditional and supervised approaches, representation learning techniques have gained traction in spectral analysis. Autoencoders and variational autoencoders (VAEs) have been used as a preprocessing step to extract compact spectral representations, facilitating parameter estimation and transfer learning across different observational datasets \citep{Yang2015autoencoder, MasBuitrago2024using, Leung2023variational, buck2024deep}. 
An advancement in this field is Disentangled Representation Learning (DRL, \citealt{Bengio2013representation}), which aims to separate independent generative factors in the data. In a disentangled representation, each latent variable is sensitive to changes in a single underlying generative factor of the data, while being invariant to changes in the others. 
This property improves explainability by ensuring that learned features align with meaningful generative factors. Additionally, it improves generalization by isolating task-relevant features from irrelevant variations in the data \citep{dittadi2020transfer, mu2021sdf, zhang2022towards, li2022ood,yoo2023disentangling}.

While DRL has been extensively explored in domains such as computer vision (\emph{e.\,g.} \citealt{wang2023disentangled}), robotics (\emph{e.\,g.} \citealt{emukpere2025disentangled}), and natural language processing (\emph{e.\,g.} \citealt{Carvalho2022learning}), its principles naturally extend to astrophysical spectra, which arise from a combination of independent physical properties—such as temperature, gravity, and individual elemental abundances—each contributing distinct features to the observed signal.
\cite{demijolla2021disentangled},\cite{santoveña2024method} successfully applied this approach to learn spectral representations that isolate variations in chemical abundances in simulated stellar spectra, while \cite{Manteiga2025disentangling} demonstrated it on data from the Gaia Radial Velocity Spectrograph (\citealt{recio2023gaia}).

\subsection{Our Contributions}
In this work, we develop a self-supervised deep learning framework for learning disentangled chemical representations from stellar spectra. Our approach is based on a VAE to encode high-dimensional spectra into a structured, low-dimensional latent space where each latent feature is associated with a specific chemical element and captures variations in its abundance.
We choose a VAE for its probabilistic framework, which regularizes the latent space and promotes smoother and more regular representations \citep{Kingma2013auto}.

Because unsupervised disentanglement is impossible without additional constraints, as multiple latent representations can explain the data equally well \citep{locatello2018challenging}, we incorporate physically motivated inductive biases to obtain meaningful, element-specific representations.
In this context, inductive biases act as a bridge between physical interpretability and model freedom: they constrain the solution space just enough to make the learned factors physically meaningful, while avoiding the need for complete and explicit supervision.

Our approach can therefore be viewed as a step towards model-free abundance inference: the model learns directly from spectra rather than relying on synthetic atmosphere models or externally fitted labels, while remaining consistent with basic spectroscopic principles. It yields latent variables that are physically interpretable, regularized, and sensitive to abundance-driven spectral variations, and provides reconstruction residuals that act as indicators of deviations from the learned chemical manifold.

With these properties, the model can serve as a tool for guiding further spectroscopic analysis with varying amounts of labeled data, such as chemical abundance estimation and anomaly detection in large-scale surveys.
In this first study, we specifically focus on learning disentangled representations of three chemically relevant dimensions--metallicity, carbon, and $\alpha$-element abundances--from synthetic spectra and show that these representations have the potential to uncover chemically enhanced or depleted stars—specifically, $\alpha$-poor, metal-poor (\alphapmp) and carbon-enhanced, metal-poor (CEMP) stars—in a self-supervised manner.

This paper is organized as follows. In Section \ref{sec:methods}, we describe the proposed methodology, including the architecture and training of our model and the subsequent search for chemically enhanced and depleted stars. Section \ref{sec:data} details the synthetic spectral dataset, and Sect. \ref{sec:results} presents our experimental results, highlighting findings on $\alpha$-poor, metal-poor and CEMP stars. Finally, Section \ref{sec:conclusions} discusses the implications of our work for understanding the Milky Way’s formation history and outlines prospects for future applications to large-scale spectroscopic datasets. 
\section{Methods}\label{sec:methods}
In this section we present our method, which consists of two main stages: (1) learning a structured representation of stellar spectra that depends only on chemistry and (2) leveraging this representation to identify chemically enhanced or depleted stars. Our approach is inspired by \cite{demijolla2021disentangled}, with key modifications to ensure disentanglement of spectral features linked to different chemical abundances.

Section \ref{sec:problem} introduces the problem formulation and the philosophy behind our approach. Sections \ref{sec:reprlearning}-\ref{sec:model}-\ref{sec:gradients} detail the deep learning framework and model architecture, emphasizing the structured latent space that enables interpretability. Finally, in Section \ref{sec:ad}, we describe how this learned representation is used to identify chemically distinct stars. While these stars may not be clearly separated in chemical space, our method enables their detection by characterizing their position relative to the overall distribution of latent variables, allowing us to highlight departures from typical abundance patterns.

\subsection{Problem definition}\label{sec:problem}
The observed spectrum $\vec{s} \in \mathbb{R}^d$ can be expressed as  
\begin{align}
    \vec{s} = f(T_{\rm eff}, \log g, \vec{A}) + \boldsymbol{\epsilon},
\end{align}
where $T_{\rm eff}$ is the effective temperature, $\log g$ is the logarithm of the surface gravity, \(\vec{A} = (A_1, A_2, ..., A_m)\) represents the chemical abundances, where $m$ is the number of elemental abundances considered, and \(\boldsymbol{\epsilon}\) accounts for observational uncertainties, measurement noise, and unmodeled spectral variations.
Because the dominant variations in $\vec{s}$ are driven by $T_{\rm eff}$ and $\log g$, stars with different chemical compositions may still have similar spectra.
We focus on $T_{\rm eff}$ and $\log g$ as the main atmospheric parameters, since additional quantities—such as microturbulence, macroturbulence, or mixing length—are artifacts introduced in 1D models to approximate convective effects, and are not required in more realistic 3D simulations where such processes are treated self-consistently \citep{magic2013stagger,magic2014staggerV,magic2015staggerII}.

We map the spectra into a lower-dimensional latent space, capturing the essential generative factors of the data:
\begin{align}
    &\vec{h} = g_{\theta}(\vec{s}),\label{eq:encoder}
\end{align}  
where $g_{\theta}$ is a function, parameterized by $\theta$, which maps the input spectrum $\vec{s}$ to a latent representation $\vec{h}=\vec{h}(T_{\rm eff}, \log g, \vec{A}) \in \mathbb{R}^{d'}$, $d' \ll d$.
Without additional constraints, the resulting latent representation is likely to be dominated by variations in \tefflogg, rather than isolating chemical information.  
If each component of $\vec{h}$ retains information about $T_{\rm eff}$ and $\log g$, chemically similar stars at different temperatures or gravities are likely to appear dissimilar in each component of the latent space.
To obtain a chemically meaningful representation, we require at least a portion $\vec{z}$ of the latent space to capture only variations in chemical abundances:
\begin{align}
    \vec{z} = \vec{z}(\vec{A})
\end{align}

Finally, we impose additional structure such that each latent variable $z_i$ corresponds directly to a single abundance $A_i$:  
\begin{equation}
    z_i \propto	 A_i, \quad \forall i=1,\dots,m.
\end{equation}  
While this formulation is general and can accommodate any number $m$ of chemical elements, in this work we focus on metallicity, carbon, and $\alpha$-element abundances. Thus, we set $m=3$ and
\begin{equation}
    \vec{z} = (z_{\mathrm{M}}, z_{\mathrm{C}}, z_{\alpha}) \propto (\mathrm{[Fe/H]},\; \mathrm{[C/Fe]},\; \mathrm{[\alpha/Fe]}).
\end{equation}

\subsection{Disentanglement from non-chemical generative factors}\label{sec:reprlearning}
To obtain a representation space that is independent of non-chemical generative factors, in this case $T_{\rm eff}$ and $\log{g}$, we employ disentangled representation learning, following an approach similar to \cite{demijolla2021disentangled}.  
Our representation learning model is based on VAEs, neural networks composed of an encoder-decoder pair.  
The encoder, which models $g_{\theta}$, maps spectra $\vec{s}$ to a lower-dimensional latent representation, $\vec{h}=g_{\theta}(\vec{s})$ (Eq.~\ref{eq:encoder}).  
We split this representation into two components:
\begin{itemize}
    \item A chemical latent space $\vec{z}$, which is designed to encode information about chemical abundances.
    \item An auxiliary latent space $\bf{w}$ which captures variations due to the non-chemical generative factors $\vec{u}=(T_{\rm eff},\log{g})$. 
\end{itemize}
A decoder $h_{\phi}$ then reconstructs the spectrum using both:
\begin{equation}
    \hat{\vec{s}} = h_{\phi}(\vec{z},\vec{w}),
\end{equation}  
Unlike previous approaches (\emph{e.\,g.} \citealt{demijolla2021disentangled}) that provide $\vec{u}$ explicitly as inputs to the decoder, this formulation designs the encoder to disentangle $\vec{u}$ from $\vec{z}$, while still allowing $\vec{u}$ to influence the encoding of the spectra.

To further encourage independence between the chemical latent representation $\vec{z}$ and the non-chemical parameters $\vec{u}$, we integrate an adversarial training mechanism inspired by Fader Networks \citep{Lample2017fader}. We introduce an auxiliary neural network $d_\psi$, often referred to as the adversary, which attempts to infer $\vec{u}$  from $\vec{z}$. Concurrently, the encoder $g_{\theta}$, is trained to minimize the discrepancy between reconstructed and original spectra, while also producing a representation $\vec{z}$ that prevents this adversary from accurately recovering $\vec{u}$. 

This adversarial setup is implemented as a minimax optimization: the adversary is optimized to minimize its prediction loss $\mathcal{L}_{\rm adv}$, while the encoder is optimized to maximize it. Once the adversary fails to extract any information about $\vec{u}$ from $\vec{z}$, we can conclude that $\vec{z}$ has become independent of non-chemical factors, ensuring a disentangled representation that primarily encodes chemical abundances. Additionally, during training, we optimize a linear rescaling $l_{l}$ of $\vec{w}$  to predict $\vec{u}$, which further aligns the auxiliary latent space with the non-chemical parameters.

Thus the goal of training is to find the parameters $\theta,\,\phi,\,\psi,\,l$  that minimize the overall objective function:
\begin{align}
    \min_{\theta, \phi, l} \max_{\psi} \mathcal{L}_{\text{rec}} + \beta \mathcal{L}_{\text{KL}} + \mathcal{L}_{\text{lin}} - \lambda \mathcal{L}_{\text{adv}},
    \label{eq:objective}
\end{align}
where $\mathcal{L}_{\text{rec}}$ is the reconstruction loss, which we choose to implement as the log-likelihood, and $\mathcal{L}_{\text{KL}}$ is the KL divergence regularization term from the VAE, encouraging the latent space to follow a Normal distribution.
In the standard setup this prior distribution is diagonal, encouraging independence across latent dimensions. Here, we extend it to a more general Gaussian prior, with a covariance matrix informed by astrophysical knowledge of abundance trends. This does not fix the correlations in the data, but rather acts as a soft inductive bias: the reconstruction loss still drives the model to recover the true abundance structure, while the prior regularizes the latent dimensions.

The adversarial loss $\mathcal{L}_{\text{adv}}$ is defined as the sparse categorical cross-entropy, a classification loss. Since the labels $\vec{u}$ are continuous and two-dimensional, we discretize them into one-dimensional bins. 
By choosing classification over regression—i.e., predicting discrete classes instead of continuous values—we simplify the task for the adversary, making the adversarial training process more stable and, consequently, improving disentanglement.

Finally, the linear loss $\mathcal{L}_{\text{lin}}$ is the mean squared error between the $l_l(\vec{w})$ and $\vec{u}$.
The definition of these losses are presented in Appendix~\ref{app:losses_metrics}.
Although multiple loss terms are involved, each contributes to compatible objectives, ensuring that the overall complexity remains manageable and aligned with the primary goal of disentangling the latent representation.

In our implementation, the coefficients controlling the relative importance of latent space regularization and disentanglement ($\beta$ and $\lambda$ in Eq. \ref{eq:objective}, respectively) are updated according to a stepwise schedule based on the training iteration (see Appendix \ref{app:training_details}).
We found that the stepwise schedule prevents the adversary from collapsing into a dummy classifier early in training, ensuring more effective disentanglement.

\subsection{The model}\label{sec:model}
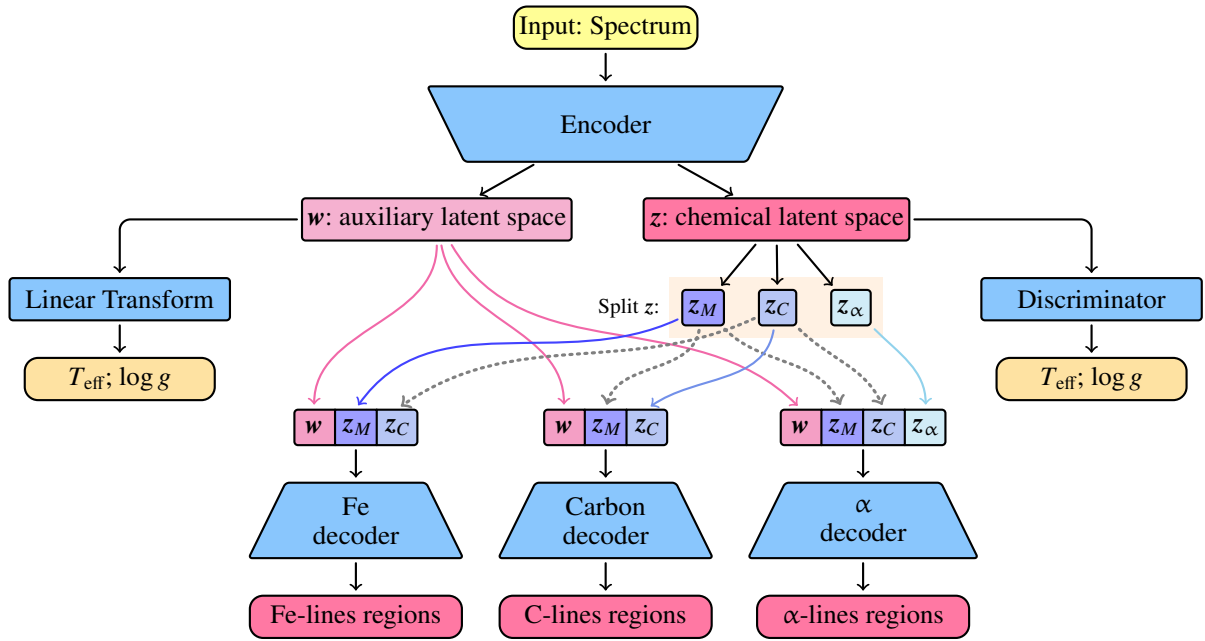
\begin{figure*}[t!]
    \centering
    \newcommand{\splitmcolor}{blue}
\newcommand{\splitccolor}{RoyalBlue}
\newcommand{\splitalphacolor}{SkyBlue}
\definecolor{customPink}{RGB}{255, 86, 140}    %
\definecolor{customBlue}{RGB}{107, 184, 252}    
\definecolor{customYellow}{RGB}{255, 254, 150}    %
\definecolor{customYellow2}{RGB}{254, 226, 162}    %

\tikzstyle{every pin edge}=[<-, shorten >=8pt, shorten <=8pt]
\tikzstyle{base node}=[rectangle, draw, thick, color=black, fill=magenta!47, inner sep=2pt, rounded corners=0.47mm, align=center, minimum height=0.56cm]

\tikzstyle{encoder node}=[base node, trapezium, trapezium angle=-64, fill=customBlue!80, minimum width=1.cm, text width=1.28cm, minimum height=1cm]
\tikzstyle{decoder node}=[encoder node, trapezium angle=64]

\tikzstyle{input node}=[base node, rounded corners=5pt, fill=orange!29]
\tikzstyle{output node}=[input node, fill=customPink!80, minimum width=2.5cm]

\tikzstyle{arrow}=[->, thick, shorten >=1.5pt, shorten <=1.5pt, rounded corners]
\tikzstyle{arrow-aux}=[arrow, color=magenta!74]
\tikzstyle{arrow-zm}=[arrow, color=\splitmcolor!74]
\tikzstyle{arrow-zc}=[arrow, color=\splitccolor!74]
\tikzstyle{arrow-za}=[arrow, color=\splitalphacolor!92]

\tikzstyle{transform node}=[base node, fill==magenta!47, text width=1.91cm, text centered]
\tikzstyle{split node}=[base node, fill=orange!47, minimum height=0.47cm]
\tikzstyle{group node}=[rectangle, line width=0.38pt, inner sep=0.15cm, fill=pink!29]

\tikzstyle{concat node}=[rectangle split, draw, thick, color=black, inner sep=2pt, rounded corners=0.47mm, align=center, minimum height=0.47cm, rectangle split horizontal, rectangle split parts=3, text width=0.38cm]

\def\horizsep{0.92cm}
\def\vertsep{0.47cm}

\begin{tikzpicture}[node distance=\vertsep and \horizsep, line join=round, line cap=round
]

\begin{scope}[local bounding box=latents-container]
\node[base node, fill=magenta!38] (aux) {$\vec{w}$: auxiliary latent space};
\node[base node, fill=customPink!80, right=of aux] (chem) {$\vec{z}$: chemical latent space};
\matrix [matrix, group node, fill=orange!11, column sep=0.47*\horizsep, label={left:\tiny Split $\vec{z}$:}, below=of chem] (split-chem) {
    \node[split node, fill=\splitmcolor!38] (zm) {$\vec{z}_M$}; &
    \node[split node, fill=\splitccolor!38] (zc) {$\vec{z}_C$}; &
    \node[split node, fill=\splitalphacolor!38] (za) {$\vec{z}_\upalpha$};\\
};
\end{scope}

\node[encoder node, above=of latents-container] (encoder) {Encoder};
\node[input node, fill=customYellow, above=of encoder] (spectrum) {Input: Spectrum};

\draw[arrow] (spectrum) -- (encoder);
\draw[arrow] (encoder) -- (aux);
\draw[arrow] (encoder) -- (chem);
\draw[arrow] (chem) -- (zm);
\draw[arrow] (chem) -- (zc);
\draw[arrow] (chem) -- (za);

\node[concat node, below=2*\vertsep of latents-container, rectangle split part fill={magenta!47,\splitmcolor!38,\splitccolor!38}] (concat-carbon) {$\vec{w}$
                   \nodepart{two} $\vec{z}_M$
                   \nodepart{three} $\vec{z}_C$};
    
\node[decoder node, below=of concat-carbon] (decoder-carbon) {Carbon decoder};
\node[output node, below=of decoder-carbon, minimum width=2.79cm] (output-carbon) {C-lines regions};

\node[decoder node, left=of decoder-carbon] (decoder-fe) {Fe decoder};
\node[concat node, above=of decoder-fe, rectangle split part fill={magenta!47,\splitmcolor!38,\splitccolor!38}] (concat-fe){$\vec{w}$
                   \nodepart{two} $\vec{z}_M$
                   \nodepart{three} $\vec{z}_C$};

\node[output node, below=of decoder-fe, minimum width=2.79cm] (output-fe) {Fe-lines regions};
\node[decoder node, right=of decoder-carbon] (decoder-alpha) {$\upalpha$\\decoder};
\node[concat node, above=of decoder-alpha, rectangle split parts=4, rectangle split part fill={magenta!47,\splitmcolor!38,\splitccolor!38,\splitalphacolor!38}] (concat-alpha){$\vec{w}$
                   \nodepart{two} $\vec{z}_M$
                   \nodepart{three} $\vec{z}_C$\nodepart{four} $\vec{z}_\upalpha$};

\node[output node, below=of decoder-alpha, minimum width=2.79cm] (output-alpha) {$\upalpha$-lines regions};




\draw[arrow-aux] (aux)  to[out=270,in=90] (concat-fe.one north);
\draw[arrow-aux] (aux) to[out=280,in=90] (concat-carbon.one north);
\draw[arrow-aux] (aux) to[out=300,in=135] (concat-alpha.one north);

\draw[arrow-zm] (zm) to[out=205,in=55]  (concat-fe.two north);
\draw[arrow-zm, dotted, gray, very thick] (zm) to[out=260,in=75] (concat-carbon.two north);
\draw[arrow-zm, dotted, gray, very thick] (zm) to[out=315,in=115] (concat-alpha.two north);

\draw[arrow-zc, dotted, gray, very thick] (zc) to[out=205,in=35] (concat-fe.three north);
\draw[arrow-zc] (zc) to[out=260,in=35] (concat-carbon.three north);
\draw[arrow-zc, dotted, gray, very thick] (zc) to[out=315,in=115] (concat-alpha.three north);

\draw[arrow-za] (za.south east) to[out=315,in=90] (concat-alpha.four north);

\draw[arrow] (concat-carbon) -- (decoder-carbon);
\draw[arrow] (decoder-carbon) -- (output-carbon);
\draw[arrow] (concat-fe) -- (decoder-fe);
\draw[arrow] (decoder-fe) -- (output-fe);
\draw[arrow] (concat-alpha) -- (decoder-alpha);
\draw[arrow] (decoder-alpha) -- (output-alpha);

\node[base node, fill=customBlue!80, below left=of aux, minimum width=2.9cm] (transf-linear) {Linear Transform};
\node[output node,fill=customYellow2, below=of transf-linear] (teff-linear) {$T_{\text{eff}}$; $\log g$};

\draw[arrow] (aux) -| (transf-linear);
\draw[arrow] (transf-linear) -- (teff-linear);

\node[base node, fill=customBlue!80, below right=of chem, minimum width=2.9cm] (discrim) {Discriminator};
\node[output node, fill=customYellow2,below=of discrim)] (teff-discrim) {$T_{\text{eff}}$; $\log g$};

\draw[arrow] (chem) -| (discrim);
\draw[arrow] (discrim) -- (teff-discrim);

\end{tikzpicture}
    \caption{Main model. A spectrum is given as input. An encoder maps it to the latent space, divided into chemical latent space $z=(z_{\rm M},z_{\rm C},z_{\rm \alpha})$, and auxiliary latent space $\vec{w}$. Three decoders map different components of $\vec{z}$, along with $\vec{w}$, to separate regions of the input spectrum. A linear transform is applied to the auxiliary space to obtain ($T_{\rm eff},\log g$). The discriminator is trained to predict the non-chemical parameters ($T_{\rm eff},logg$) from $\vec{z}$, while the encoder is trained to make this impossible for the discriminator to predict these parameters from $\vec{z}$ alone. Grey dotted arrows represent the absence of gradient propagation (see Sect.~\ref{sec:gradients}). At inference time the decoders, the discriminator and the linear transform module are discarded.}
\label{fig:model}
\end{figure*}
To link each latent feature to specific chemical abundances, we develop a structured encoder-decoder architecture where different decoders specialize in reconstructing spectral regions dominated by particular chemical elements.

As we are interested in three chemical abundances, our model, illustrated in Fig. \ref{fig:model}, consists of a single encoder and three decoders, each responsible for reconstructing different spectral regions:
\begin{enumerate}
    \item The metal-decoder receives two latent features: $z_{\rm M}$, which captures general metallicity and $z_{\rm C}$, shared with the carbon-decoder (described below). 
    It reconstructs spectral regions dominated by iron while also taking $\vec{w}$ as input.
    \item The carbon-decoder shares $z_{\rm M}$ with the metal-decoder but has an additional feature, $z_{\rm C}$, specific to carbon. It reconstructs spectral regions dominated by carbon while also taking $\vec{w}$ as input.
    \item The $\alpha$-decoder builds upon the representations learned by the other decoders: it shares $z_{\rm M}$ with the metal-decoder, $z_{\rm C}$ with the carbon-decoder, and incorporates a separate feature $z_{\rm \alpha}$, dedicated to $\alpha$-elements.
    It reconstructs spectral regions dominated by $\alpha$-element lines while also taking $\vec{w}$ as input.
\end{enumerate}
The shared use of $z_{\rm M}$ and $z_{\rm C}$ across decoders ensures that each decoder can still access relevant information needed to reconstruct its assigned spectral region, even when it includes contributions from multiple absorbers.

This design is modular and readily extensible to accommodate any number of chemical components by introducing additional decoders specialized to reconstruct different spectral regions.
The specific spectral regions—whether broad bands, individual absorption lines, or other wavelength intervals—assigned to each decoder will depend on the resolution, wavelength coverage, and other properties of the instrument.

Our autoencoder follows the architecture proposed by \cite{melchior2023autoencoding} which employs a convolutional encoder and a fully connected decoder for galaxy spectra.
The encoder uses a PReLU activation function \citep{he2015delving}, while the decoders use an activation function of the form
\begin{equation}
a(x)=\left[\gamma+\frac{1-\gamma}{\left(1+\mathrm{e}^{-\beta\odot x}\right)}\right]\odot x,
\end{equation}
where $\gamma$ and $\beta$ are trainable parameters \citep{Alsing2020SPECULATOR}. 
This activation function allows the model to handle smooth features for small $\beta$ and sharp changes in gradients as $\beta \to \infty$, making it particularly effective for spectral line modeling.
This type of network and activation function has also been utilized in \cite{buck2024deep}, showing also its suitability for stellar spectra.

The adversary is a feed-forward neural network.
Specifics are reported in Appendix \ref{app:training_details}.
\subsection{Training with selective gradient flow}\label{sec:gradients}
In standard training, the reconstruction loss from each decoder is typically backpropagated through the entire latent space. This means, for example, that the $\alpha$-decoder can update not only the latent feature associated with $\alpha$-elements, but also those associated with iron or carbon. As a result, latent features may become entangled, with each one capturing mixed information from multiple chemical factors. This is especially problematic in stellar spectra, where strong features like metallicity can dominate the learning signal and suppress weaker patterns, such as those associated with $\alpha$-elements.

To address this, we implement a selective gradient flow strategy, restricting the propagation of reconstruction losses to their relevant latent dimensions. Specifically:
\begin{itemize}
    \item The $\alpha$-decoder backpropagates gradients only to the latent feature associated with $\alpha$-elements ($z_{\rm \alpha}$).
    \item The carbon-decoder backpropagates gradients only to the latent feature associated with carbon ($z_{\rm C}$).
    \item The metallicity (iron) latent feature ($z_{\rm M}$) is used by all decoders but remains unaffected by the reconstruction losses from the $\alpha$ and carbon decoders.
\end{itemize}

This means that, the loss from the $\alpha$-decoder influences only the latent feature for $\alpha$-elements, the carbon-decoder’s loss affects only the latent feature associated with carbon, and the metal-decoder’s loss affects only the latent feature associated with metallicity. This prevents the reconstruction losses for each region from affecting the latent features associated with the other regions. Put differently, each decoder can still leverage information from other latent features (e.g., the $\alpha$-decoder still uses $z_{\rm M}$ and $z_{\rm C}$ to account for metallicity and carbon contributions in its spectral region), but it cannot affect them.

By maintaining this selective gradient flow, we ensure that each latent feature distinctly represents a specific chemical factor. 
Without this control, the strong metallicity features in stellar spectra would dominate the latent space, making it difficult for the model to learn representations for other chemical abundances (see Appendix~\ref{app:gradflow} for a comparison of results with and without selective gradient flow).
\subsection{Identifying chemically enhanced/depled stars}\label{sec:ad}
After training, the latent space is chemically structured by design: each \textit{chemical} latent feature is aligned with a specific chemical abundance -- in our case one with ${[\rm Fe}/{\rm H}]$, another with $[\alpha/{\rm Fe}]$, and a third with $[{\rm C}/{\rm Fe}]$. Moreover, thanks to the KL divergence regularization term (\ref{eq:KL}), the latent features follow approximately Normal distributions (mean $\mu=0$ and standard deviation $\sigma=1$). This means that uncommon stars appear in the tails of the distribution and we can define a star as enhanced or depleted if it lies beyond a chosen threshold $k$ of standard deviations in any of the latent features. This statistically principled approach directly exploits the Gaussian nature of the latent space, offering an efficient and interpretable method for flagging stars with arbitrary levels of rarity.

To assess these capabilities, we focus on two categories of chemically interesting stars.
We evaluate our model on the quality of the flagging of $\alpha$-poor, metal-poor stars (\alphapmp) and carbon-enhanced, metal-poor stars (CEMP). 
These stars, rather than forming separate, isolated clusters in chemical space, are distributed within a continuous space, with approximate boundaries.
\alphapmp stars can be defined as stars with $[\alpha/{\rm Fe}]<0.0$ and $[{\rm Fe}/{\rm H}]<-1.0$ \citep[\emph{e.\,g.}][]{Nissen1997chemical,Ramirez2012oxigen, Hawkins2014relative,li2022four}, CEMP stars as metal-poor stars with $[{\rm C}/{\rm Fe}]>0.7$ \citep[\emph{e.\,g.}][]{Aoki2007carbon} and $[{\rm Fe}/{\rm H}]<-1.0$.

We transform these cuts into the latent space by computing their locations in terms of standard deviations from the expected chemical distribution's mean. In this representation, \alphapmp stars are identified as those lying beyond the chosen $k\sigma$ thresholds in both $z_{\alpha}$ (low values) and $z_{\rm M}$ (low values), while carbon-enhanced, metal-poor stars correspond to high $z_{\rm C}$ and low $z_{\rm M}$.
\section{Simulated dataset}\label{sec:data}
To generate our dataset, we synthesize spectra using iSpec \citep{blancocuaresma2014determining,blancocuaresma2019modern}, which wraps the Turbospectrum spectral synthesis code \citep{plez2012turbospectrum}, with MARCS atmospheric models \citep{Gustafsson2008marcs}. Atomic data were taken from the VALD linelist \citep{Piskunov1995vald}.

We created a set of five simulated catalogs of 40\,000 low-resolution ($R=2200$) spectra each, covering a wavelength range of $[4\,000, 10\,400]Å$, sampled with a logarithmic step of $\Delta\log_{10}(\lambda/\AA) = 0.0001$. 
These values were chosen to roughly reflect typical characteristics of spectra from modern large-scale surveys of the Galactic Halo, providing a realistic yet flexible setup for testing our approach. 
The spectral regions we selected for each decoder, based on element-specific features within this resolution and wavelength range, are detailed in Appendix~\ref{app:regions}.

Since we are interested in three chemical abundances, we synthesize spectra that depend on five parameters: effective temperature ($T_{\rm eff}$), surface gravity ($\log{g}$), metallicity ([${\rm Fe}/{\rm H}$]), $\alpha$-elements abundance ([$\alpha/{\rm Fe}$]), and carbon abundance ([${\rm C}/{\rm Fe}$]).
For each spectrum, $T_{\rm eff}$ and $\log{g}$ were generated by sampling from uniform distributions with range $[3000,7000]$~K and $[2.5,6]$~dex, respectively. 
While these distributions do not match the usual parameter spaces covered by stellar surveys, they provide a wider range of variability, increasing the complexity of the training set.

\begin{figure*}[tb]
    \centering
    \includegraphics[width=\linewidth]{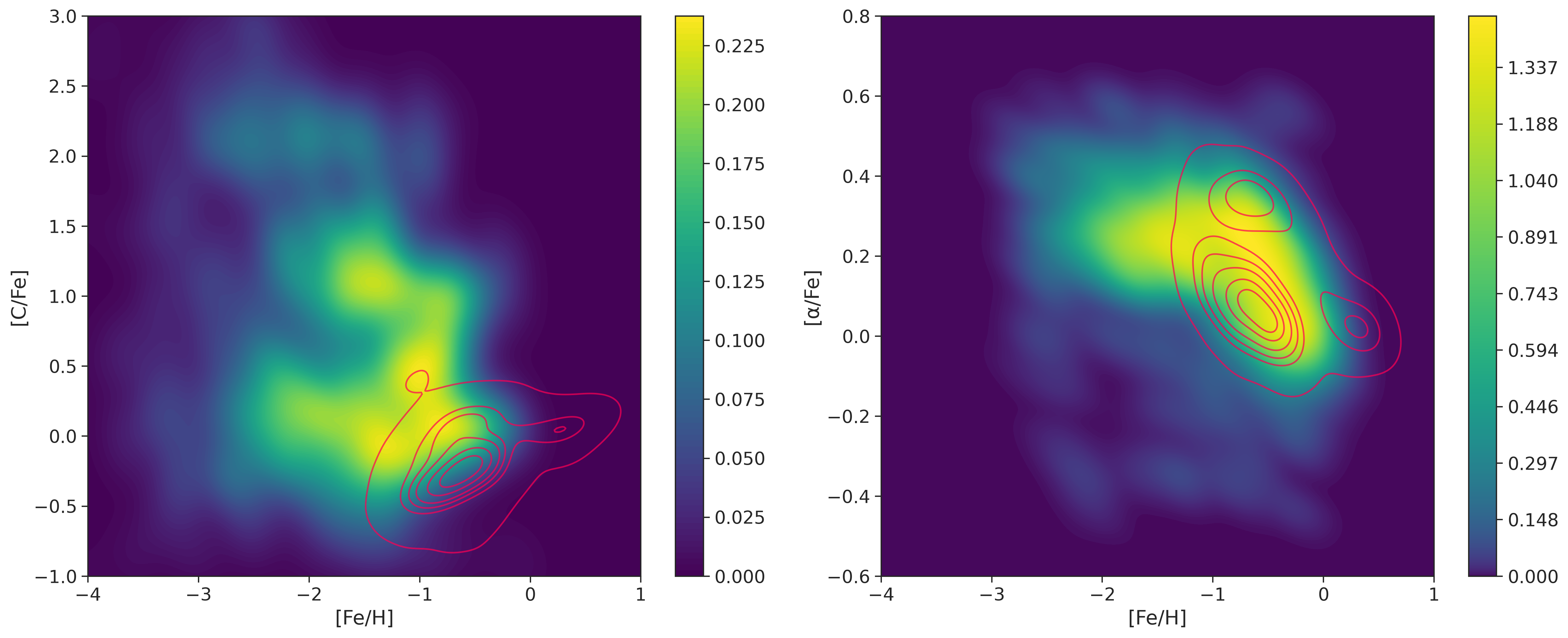}
    \caption{Probability distribution functions (PDFs) for the chemical abundances used to sample the stellar properties in our dataset,with lighter colors indicating higher probability densities. \textbf{Left panel}: PDF of ${\rm  [C}/{\rm Fe]}$ versus ${\rm  [Fe}/{\rm H]}$. \textbf{Right panel}: PDF of ${\rm  [\alpha}/{\rm Fe]}$ versus ${\rm  [Fe}/{\rm H]}$. In both panels, the pink contours show the abundance distributions of halo stars from the APOGEE survey, as provided by the astroNN catalog. APOGEE does not reach the lowest metallicities present in our simulated dataset, which explains the differences at low ${\rm  [Fe}/{\rm H]}$.}
    \label{fig:distro}
\end{figure*}

The chemical abundances are sampled from the joint distribution
\begin{align}
    r({\rm  [Fe}/{\rm H]},&{\rm  [C}/{\rm Fe]},{\rm  [\alpha}/{\rm Fe]})=\\ &\frac{1}{Z}\,p({\rm  [Fe}/{\rm H]},{\rm  [C}/{\rm Fe]})\times q({\rm  [Fe}/{\rm H]},{\rm  [\alpha}/{\rm Fe]}),
\end{align}
where $p$ and $q$ are the probability distribution functions for the pairs $({\rm  [Fe}/{\rm H]},{\rm  [C}/{\rm Fe]})$ and $({\rm  [Fe}/{\rm H]},{\rm  [\alpha}/{\rm Fe]})$, respectively, observed in the Galactic Halo population (e.g. \citealt{ArdernArentsen2025predicting,Chiappini2001formation}, see Fig. \ref{fig:distro}). $Z$ is a normalization constant
\begin{align}
Z = \int p\times q \; d\mathrm{[Fe/H]}\, d\mathrm{[C/Fe]}\, d\mathrm{[\alpha/Fe]}.
\end{align}

In this distribution we find:
\begin{align}
    \mu_{\rm [Fe/H]} &= -1.23,\quad \sigma_{\rm [Fe/H]} = 0.57 \\
    \mu_{\rm [C/Fe]} &= 0.62 \quad \sigma_{\rm [C/Fe]} = 0.75 \\
    \mu_{\rm [\alpha/Fe]} &= 0.22 \quad \sigma_{\rm [\alpha/Fe]} = 0.15
\end{align}
To better mimic realistic observations, we add Gaussian noise to the continuum-normalized spectra, with mean zero and a standard deviation of $\sigma=1/{\rm SNR}$, where the signal-to-noise ratio (SNR) is drawn from a normal distribution with a mean of 50 and a standard deviation of 10. In addition to spectral noise, we perturb \tefflogg by applying a $2\%$ relative random error, to train the disentanglement mechanism under uncertain stellar parameters.

\section{Results}\label{sec:results}
We implement the architecture introduced in Sect. \ref{sec:model} with PyTorch \citep{Paszke2019pytorch} and train it for 4000 epochs using the Adam optimizer \citep{kingma2014adam} and the 1Cycle learning rate schedule \citep{smith2019super} with a maximum learning rate of $10^{-3}$.
The model is trained and evaluated separately on the five simulated catalogs described in Sect. \ref{sec:data}, using 70\% of the spectra for training and the remaining 30\% for evaluation.
On an NVIDIA RTX A4000 GPU, training takes approximately six hours.
Unless otherwise specified, reported results are either representative samples or averages over different simulated catalogs.

\subsection{Evaluating reconstruction capabilities}
\begin{figure*}
    \centering
    \includegraphics[width=\linewidth]{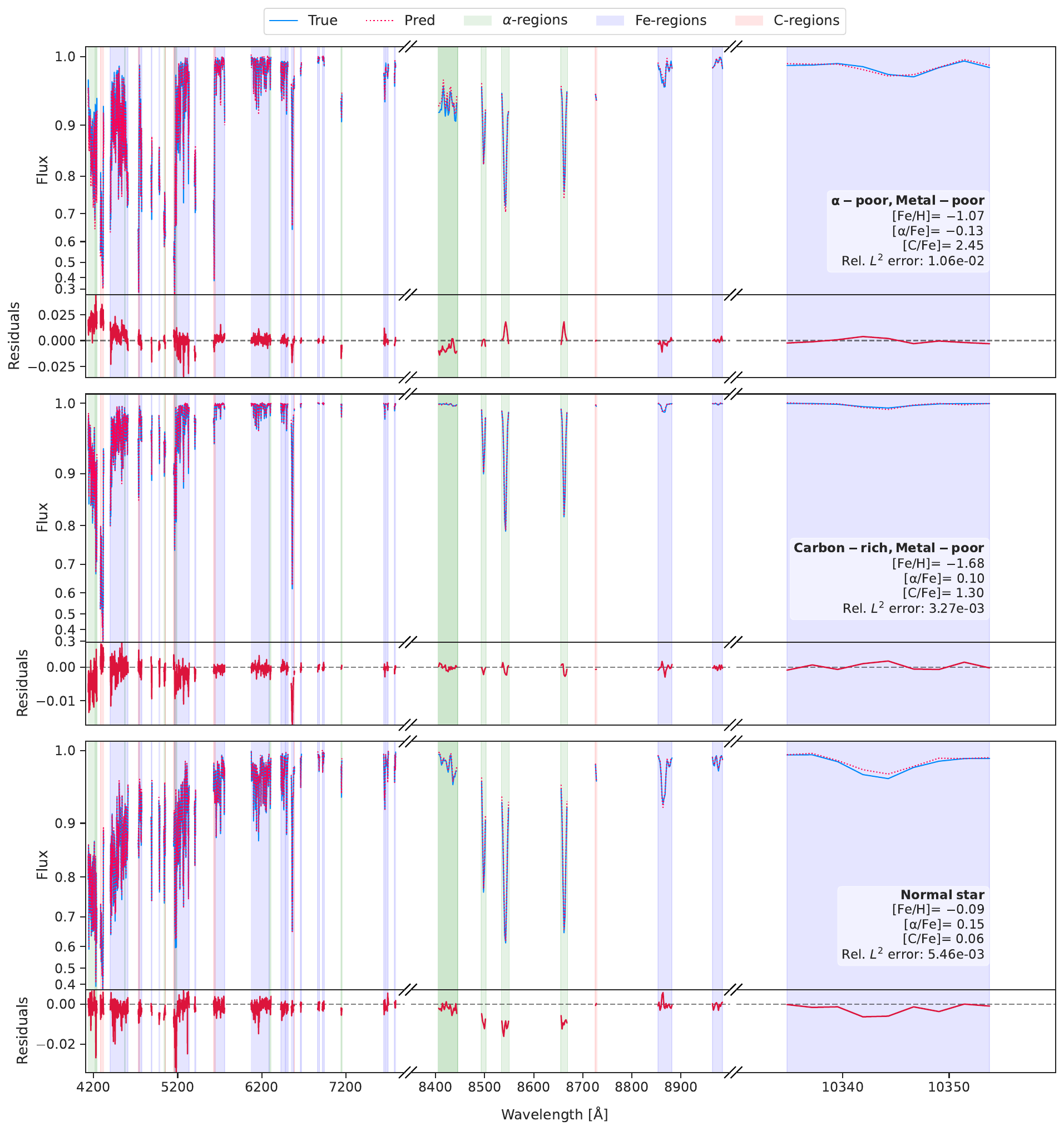}
    \caption{Comparison of original (before noise perturbation, blue) and reconstructed spectra (pink) for three chemical types: an $\alpha$-poor, metal-poor star (top row), a carbon-rich, metal-poor star (middle row), and a solar-like star (bottom row). The residuals (reconstructed minus original) are shown in red in the bottom subpanels. Shaded regions highlight spectral domains handled by the different decoders, as indicated in the legend.}
    \label{fig:reconstruction}
\end{figure*}

To assess reconstruction quality, we compare original and reconstructed spectra. Figure \ref{fig:reconstruction} presents representative examples, illustrating the agreement between input and output.

The average relative $L^2$ error (Eq. \ref{eq:rel_l2}) for all the samples in the validation set is $0.006$. For $\alpha$-poor, metal-poor stars, the error is $0.01$, for carbon-rich, $0.005$, and for solar-like composition stars, $0.004$. These results align with expectations, as the model training is focused more in reconstructing the most common stars in the dataset. In particular, only $\approx 10\%$ are $\alpha$-poor, metal-poor stars.
Overall, the model achieves a reconstruction error below the noise level in the spectra. This implies that the latent space effectively captures the essential spectral information.

\subsection{Latent space as a representation of chemical composition}\label{sec:chemical_z}
The central goal of our architecture is to learn a latent space that captures only chemical information, disentangled from non-chemical physical parameters such as effective temperature and surface gravity, $\vec{u} = (T_{\mathrm{eff}}, \log g)$, with each feature aligned to a specific chemical element.
This disentanglement and alignment with the intended chemical abundances is quantified in Table~\ref{table:correlations}, in terms of the average absolute Pearson correlation coefficients ($r$) between latent features and both chemical abundances and atmospheric parameters. Each latent feature shows strong correlation with its intended chemical abundance and negligible correlation with $T_{\mathrm{eff}}$ and $\log g$. 
Cross-correlations between latent features and other chemical abundances reflect intrinsic correlations in the data set, suggesting that the model preserves the natural structure of chemical abundances.

To test for potential multivariate nonlinear dependencies between the latent space and the atmospheric parameters, we apply a regressor (XGBoost, \citealt{Chen2016xgboost}) to predict $\vec{u}$ from the learned chemical latent features $\vec{z}$.
The regressor performs no better than a dummy baseline that simply predicts the mean of the training data, suggesting that the latent space contains no meaningful information about these physical parameters. This confirms that the model has successfully isolated chemical information in the latent space, leaving $T_{\mathrm{eff}}$ and $\log g$ disentangled.

\begin{figure*}
    \centering
    \includegraphics[width=\textwidth]{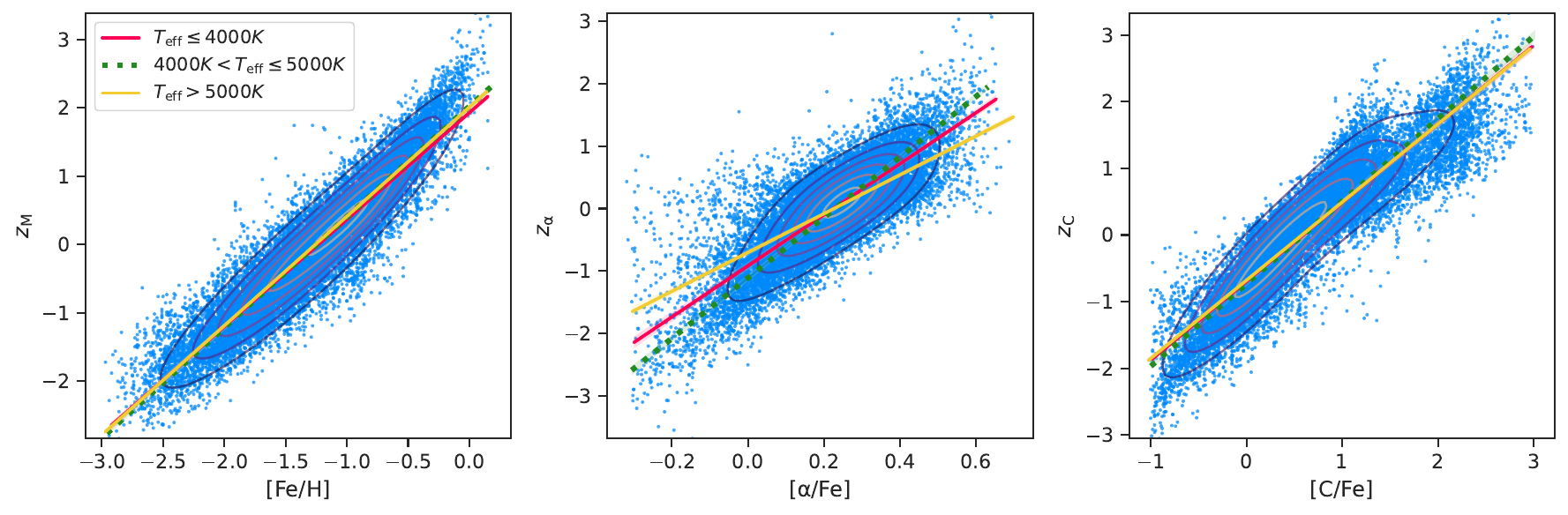}
    \caption{Contour plot of latent features (from left to right, $z_{\mathrm M}, z_{\mathrm \alpha}, z_{\mathrm C}$) and their corresponding chemical abundances. The scatter points represent individual data points, and the contour lines represent data density, with lighter contours indicating regions of higher density. Straight lines show linear fits to the latent-abundance relations for stars in three $T_{\rm eff}$ bins, as indicated in the legend.}
    \label{fig:latent_corr}
\end{figure*}

We plot each latent feature against its corresponding abundance in Fig.~\ref{fig:latent_corr}. 
These plots further demonstrate how $z_{\rm M}$ captures metallicity, $z_{\rm \alpha}$ captures $\alpha$ abundance, and $z_{\rm C}$ captures carbon abundance, effectively encoding distinct chemical properties in each feature as a rescaled representation.
When splitting the sample into low, medium, and high $T_{\rm eff}$ bins, the latent–abundance relations for metallicity and carbon remain nearly unchanged, while the $\alpha$-latent shows a modest slope variation. This is consistent with the weaker imprint of $\alpha$-elements in stellar spectra, which makes their latent representation more susceptible to residual dependence on stellar parameters.

\begin{table}[tb]
\centering       
\caption{Correlations of the latent features with atmospheric parameters.}
\begin{tabular}{lcccc}          
\toprule          
Target feature & \textbf{$z_{\rm M}$} & \textbf{$z_{\rm C}$} & \textbf{$z_{\rm \alpha}$}\\
\midrule
$\rm [Fe/H]$ & \corrFE{}* & $0.05\pm0.08$ & $0.30\pm0.11$ \\
$\rm [C/Fe]$ & $0.33\pm0.08$ & \corrC{}* & $0.09\pm0.03$ \\
$\rm [\alpha/Fe]$ & $0.33\pm0.07$ & $0.09\pm0.03$ & \corra{}* \\

$T_{\rm eff}$ & $0.03\pm0.01$ & $0.01\pm0.01$ & $0.03\pm0.02$ \\
$\log{g}$ & $0.03\pm0.01$ & $0.01\pm0.01$ & $0.03\pm0.02$ \\
\bottomrule
\end{tabular}
\tablefoot{Average and standard deviation values of the absolute Pearson correlation coefficients $r$ between each latent feature and stellar parameters, computed over the test sets. * highlights the correlation between each chemical abundance and its associated latent feature. Low correlations with \tefflogg indicate successful disentanglement from non-chemical parameters. The strongest correlations align each latent feature with its intended chemical abundance, confirming effective specialization in the latent space.}
\label{table:correlations}
\end{table}
As a sanity check, we also tested how well the learned latent features can be rescaled into abundances and compared them to a traditional supervised learning algorithm. Since the comparison involves different levels of prior knowledge and optimization, we defer the details to Appendix \ref{app:inference}.

\subsection{Identifying chemically enhanced and depleted stars in the latent space}
Since our latent space successfully disentangles chemical information (Sect.~\ref{sec:chemical_z}), we now apply this learned representation to identify chemically peculiar stars. By leveraging the model’s ability to separate chemical features in the latent space, we can systematically search for outliers in an unsupervised manner.

As described in Sect. \ref{sec:methods}, we evaluate our model based on its ability to flag two categories of interesting stars: \alphapmp stars, and CEMP stars. These classes do not occupy distinct, isolated regions in chemical space but are instead defined by approximate thresholds in elemental abundances:
\alphapmp stars are identified as those with $[\alpha/{\rm Fe}] < 0.0$ and $[{\rm Fe}/{\rm H}] < -1.0$; CEMP stars satisfy both conditions: $[{\rm C}/{\rm Fe}] > 0.7$ and $[{\rm Fe}/{\rm H}] < -1.0$.

To assess the significance of these cuts, we determine their locations in terms of standard deviations from the mean of the dataset's chemical distributions. The $[{\rm Fe}/{\rm H}]$ threshold for MP stars corresponds to $k = 0.4$ standard deviations from the mean, the $\alpha$ cutoff for $\alpha$-poor stars is at $k = 1.5$, and the $[{\rm C}/{\rm Fe}]$ threshold for CE stars is at $k = 0.1$.

In our approach, we then flag as enhanced/depleted any star that is located more than $k$ standard deviations from the mean in any of the three latent features. 
Assuming an approximately Gaussian distribution in the latent space, this thresholding approach provides a simple yet effective method to identify rare cases.

Because the relationship between latent features and chemical abundances is not explicitly constrained to be positive, we must first establish the sign of correlation before applying thresholds. This can be done using a small set of labeled samples or by visually inspecting spectra—for example, by comparing stars from different regions of the latent space and verifying which ones have higher metallicity, $\alpha$ abundance, or carbon content. Once the sign of the correlation is established, we flag as chemically enhanced or depleted any star with a latent feature value exceeding $k$ standard deviations from the mean in the appropriate direction. Table \ref{table:anomaly_detection} summarizes the precision and recall (Eqs. \ref{eq:precision}, \ref{eq:recall}) for \alphapmp and CEMP stars.
We remind the reader that these values are tied to the specific characteristics of our data set and may vary in real-world scenarios, potentially improving with higher signal-to-noise ratios or spectral resolution.

For \alphapmp stars, the model achieves a high precision of \aPMPprecision, meaning that nearly all flagged stars truly belong to this category. While the recall is lower at \aPMPrecall, this suggests that the model is highly selective, prioritizing confidence over completeness. This is not necessarily a drawback, as it ensures that the flagged stars are indeed \alphapmp, making the method particularly useful for identifying the most chemically distinct cases. The lower recall may also reflect the challenges of estimating $\alpha$-elements abundances, especially in low-resolution spectra, where individual $\alpha$-elements absorption features can be blended or difficult to distinguish.

\begin{figure*}
    \centering
    \includegraphics[width=\textwidth]{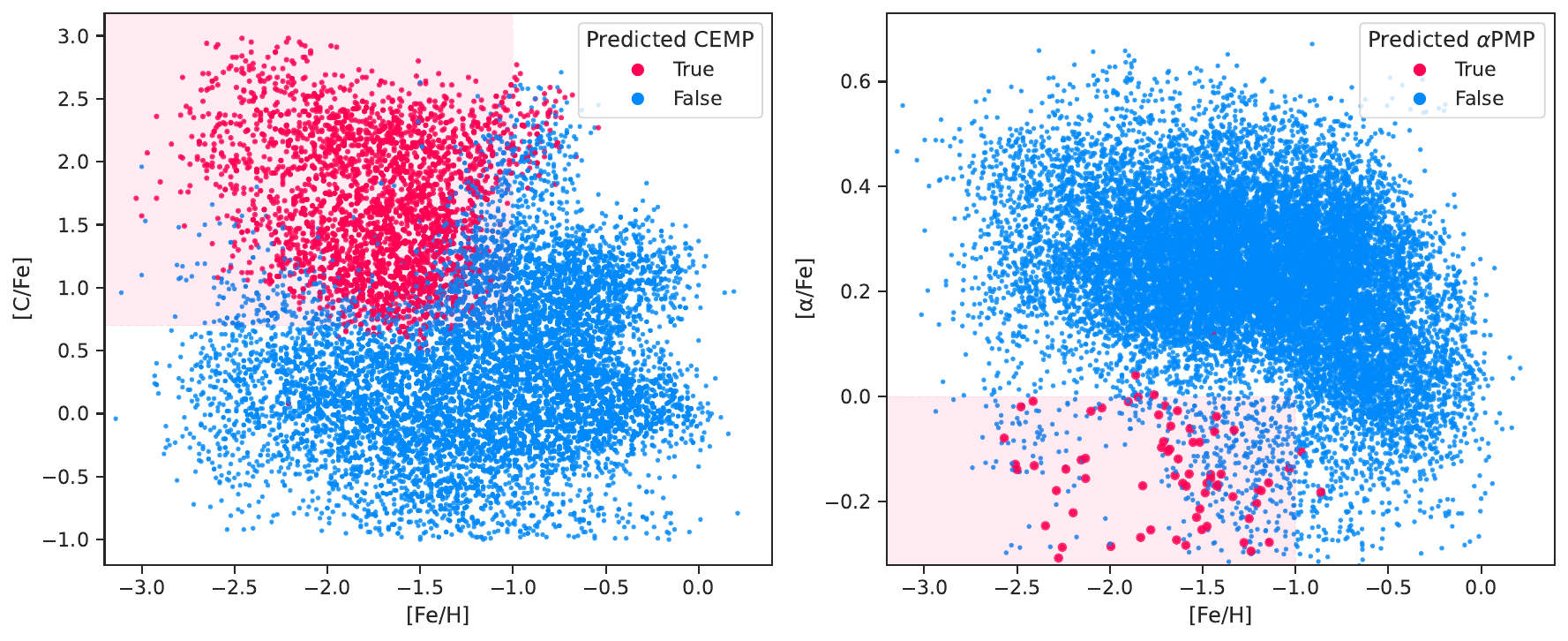}
    \caption{Distribution of stars in the true chemical space. \textbf{Left panel}: $[\mathrm{C}/\mathrm{Fe}]$ vs $[\mathrm{Fe}/\mathrm{H}]$. \textbf{Right panel}: $[\alpha/\mathrm{Fe}]$ vs $[\mathrm{Fe}/\mathrm{H}]$. The color represents the predicted class, with blue indicating stars not flagged as enhanced/depleted, and pink indicating those flagged as enhanced/depleted. The pink-highlighted regions indicate the approximate boundaries for \alphapmp and CEMP stars, as described in Sect. \ref{sec:ad}, where \alphapmp stars are defined by $[\alpha/{\rm Fe}]<0.0$ and $[{\rm Fe}/{\rm H}]<-1.0$, and CEMP stars by $[{\rm C}/{\rm Fe}]>0.7$ and $[{\rm Fe}/{\rm H}]<-1.0$.}
    \label{fig:anomalies}
\end{figure*}

For CEMP stars, the precision is similarly high at \CEMPprecision, while the recall is substantially better at \CEMPrecall. This indicates that the model captures a larger fraction of true CEMP stars while still maintaining a low false-positive rate. The higher recall for CEMP stars suggests that they are more clearly separated in the latent space, potentially due to stronger spectral features associated with carbon enhancements.
Figure \ref{fig:anomalies} illustrates the distribution of stars in chemical space, color-coded by their predicted class.

Overall, these results demonstrate that our approach successfully identifies chemically enhanced or depleted stars with high confidence, making it a valuable tool for selecting candidates for follow-up analyses of uncommon abundance patterns. The high precision across both categories confirms that the flagged stars are true positives, while the differences in recall highlight the varying challenges in identifying distinct chemical signatures.
\begin{table}[tb]
    \centering       
    \caption{Precision and recall for detection of chemically peculiar stars.}
    \begin{tabular}{lcc}          
    \toprule          
    \textbf{Target category} & \textbf{Precision} & \textbf{Recall}\\
    \midrule
    \alphapmp & \aPMPprecision & \aPMPrecall \\
    CEMP & \CEMPprecision & \CEMPrecall \\
    \bottomrule
    \end{tabular}
    \tablefoot{Precision and recall for $\alpha$-poor metal-poor stars (\alphapmp) and carbon-enhanced metal-poor stars (CEMP). Uncertainties represent standard deviations over multiple random realizations of the test set.}
    \label{table:anomaly_detection}
\end{table}
\subsection{Robustness to noise and sensitivity to perturbations}\label{sec:snr}
We now test the robustness of the learned chemical latent space to varying SNRs by examining whether noise in the input spectra affects the chemical content encoded by each latent feature. For each latent dimension, we perform a linear regression against its corresponding chemical abundance (\emph{e.\,g.}, $z_{\rm M} \rightarrow \tilde{[{\rm Fe/H}]}\approx [{\rm Fe/H}]$) and compute the residuals. These residuals quantify deviations from the expected chemical abundance based on the learned latent representation. Note that this is not intended as an abundance estimation task but rather serves as a proxy to check for potential deviations from the expected linear correlation due to noise in the spectra.

\begin{figure*}
    \centering
    \includegraphics[width=\linewidth]{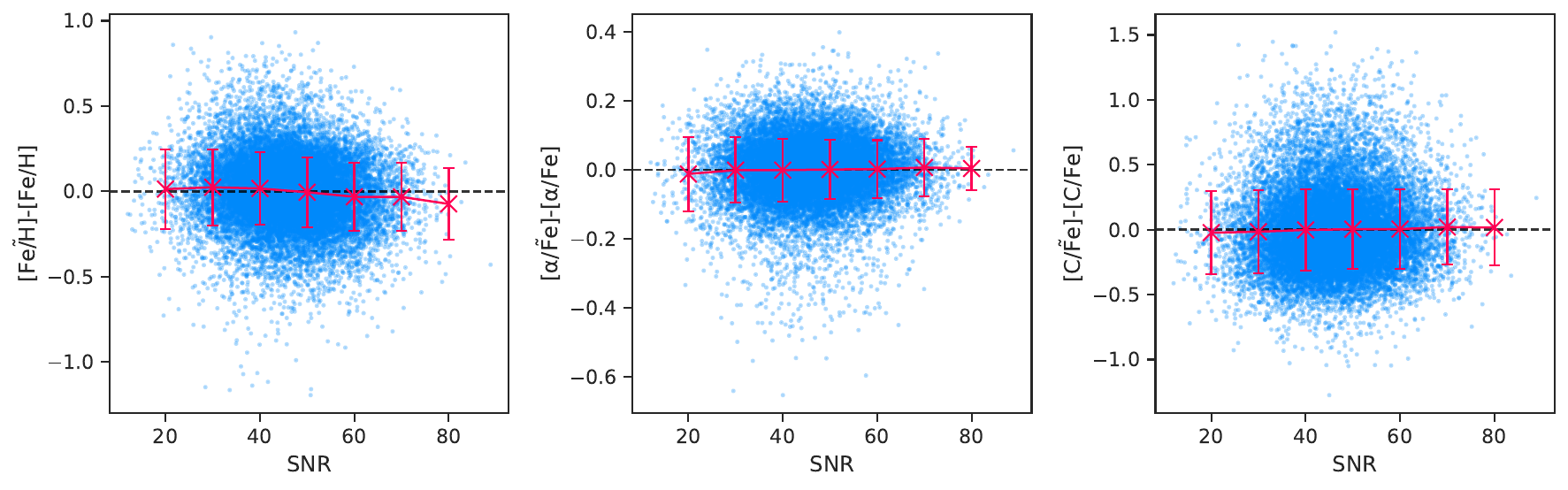}
    \caption{
    Residuals $\tilde{[{\rm X}]} - [{\rm X}]$, defined as the difference between the abundance predicted via linear regression from the latent features ($z_{\mathrm M}, z_{\alpha}, z_{\mathrm C}$) and the true values, shown as a function of SNR. 
    From left to right, the panels correspond to [Fe/H], [$\alpha$/Fe], and [C/Fe]. 
    Red markers indicate the mean residuals in bins of SNR (with width 10, ranging from 15 to 85), and the error bars show the standard deviation within each bin.}
    \label{fig:error_vs_snr}
\end{figure*}

We then plot the residuals as a function of SNR to identify any potential trends of prediction quality with respect to noise in Fig.~\ref{fig:error_vs_snr}. 
A flat residual distribution would indicate that the encoding of chemical information is stable and unaffected by the noise level of the input spectrum. Indeed, we observe no significant correlation between the residuals and SNR, confirming that the quality of the latent representation remains robust across a wide range of spectral noise conditions. This demonstrates that our model captures chemically informative features whose encoding is largely unaffected by observational noise.
We further verified this behavior by training the model on data sets with both high and low SNR, and evaluating the reconstruction error separately for each regime as a function of the training set size, as detailed in Appendix~\ref{app:noise}.

\begin{figure*}
    \centering
    \includegraphics[width=\textwidth]{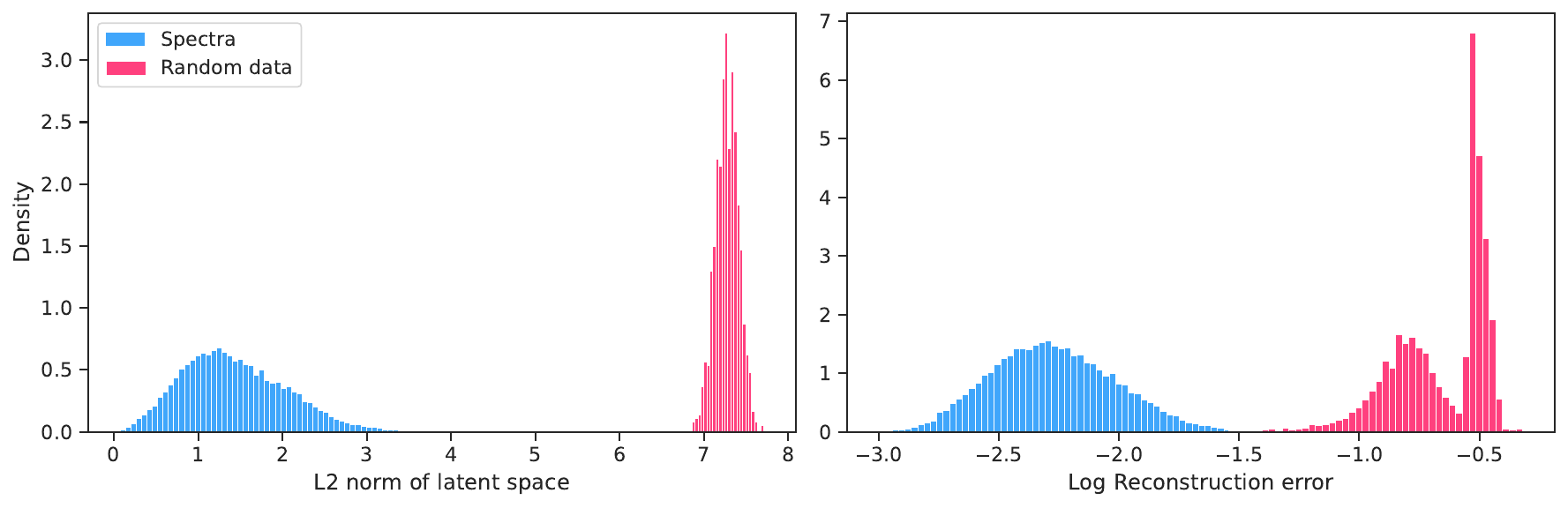}
    \caption{Distribution of the Euclidean norm of the latent representation for stellar spectra (blue) and anomalous data (pink). The clear separation between these distributions indicates that the network distinguishes true spectral features from noise.}
    \label{fig:norm_distribution}
\end{figure*}

We also explore the model's sensitivity to more extreme cases of spectral degradation. In particular, we assess its ability to handle both heavy random perturbations and systematic distortions that might occur in practice due to artifacts or data corruption.
To this end, we apply random Gaussian noise (with a standard deviation of 30\%) and introduce severe modifications, such as inserting high-amplitude spikes at random wavelengths or replacing short spectral segments ($\approx$100 pixels) with zeros or ones—thereby generating anomalous spectra. These perturbed spectra are used to evaluate the model's response to such distortions and its ability to distinguish true anomalies from ``normal'' data.
We compute the Euclidean norm of the latent chemical representation and the reconstruction error for both stellar and anomalous spectra. In Fig. \ref{fig:norm_distribution}, we show the distributions of these values for both types of input data (normal spectra vs anomalous data). We obtain a clear separation between the two distributions, with anomalous spectra exhibiting significantly larger reconstruction error and norm.

The reconstruction error, in particular, serves as a reliable metric for detecting and removing real, global anomalies, a strategy commonly used in anomaly detection. Since anomalous samples, which deviate in overall spectral properties rather than just in chemistry, are by definition rare, the model cannot learn to represent them well, resulting in poor reconstruction and thus higher errors. This makes the reconstruction error a highly effective way to identify and remove such outliers from the dataset \citep[\emph{e.\,g.},][]{sanchezsaez2021searching,holly2022autoencoder,Angiulli2023reconstruction}.

As an experiment on real data, we trained our model on a subset of LAMOST DR10 low-resolution spectra. The details of this experiment and the results are reported in Appendix~\ref{app:lamost}.
\section{Conclusions and outlook}\label{sec:conclusions}
We set out to develop a data-driven approach focused on disentangling chemical abundances that minimizes reliance on model-based assumptions and instead operates as close as possible to the observed data space. To this end, we have introduced a neural network architecture for modeling stellar spectra using a differentiable, data-driven, and physics-constrained approach. Rather than relying on a classical autoencoder, our design employs an encoder–multiple decoders architecture that explicitly disentangles spectral features associated with different chemical abundances from those driven by non-chemical atmospheric parameters. By structuring the latent space in this manner, each latent variable corresponds to a chemically meaningful component of the stellar atmosphere, improving interpretability and ensuring that the representations are independent of non-chemical factors.

Tests on simulations reveal that our architecture enables precise reconstruction of spectra, even for chemically peculiar stars. Each decoder specializes in reconstructing spectral regions dominated by specific elements (iron, $\alpha$-elements, and carbon), while targeted gradient control prevents dominant metallicity and carbon features from overshadowing the weaker $\alpha$-element signatures. The resulting latent space is highly structured and interpretable, with individual features strongly correlated with $\rm [Fe/H]$, ($r = \corrFE$), $\rm [C/Fe]$ ($r = \corrC$), and $\rm [\alpha/Fe]$ ($r = \corra$).
To further quantify the degree to which chemical abundances are encoded in the latent space, we trained simple linear regressors to map from latent features back to abundance estimates. The resulting predictions exhibit low residuals, with RMSEs of $0.21$ dex for $\rm [Fe/H]$, $0.08$ dex for $\rm [\alpha/Fe]$, and $0.31$ dex for $\rm [C/Fe]$.

Moreover, the imposed Gaussian prior on the chemical latent space facilitates a straightforward framework that can flag stars with unusual chemical compositions.
For \alphapmp stars, the model achieves a high precision of \aPMPprecision and a recall of \aPMPrecall. While the recall can be considered modest, the high precision suggests the model is selective and robust against false positives. For carbon-enhanced metal-poor (CEMP) stars, the model performs even better, with a precision of \CEMPprecision and a recall of \CEMPrecall. This demonstrates that the model can reliably identify chemically enhanced or depleted stars, with higher recall for CEMP stars suggesting that carbon enhancements are more clearly separated in the latent space.

We examined how varying noise levels in the spectra affect the structure of the latent space. By analyzing deviations from linear correlations between latent features and chemical abundances across different SNR regimes, we found that the learned representations remain robust, with no significant distortions even at low SNR.
Overall, these results demonstrate that our approach is not only capable of reconstructing stellar spectra accurately but also provides an easy to implement tool for detecting chemically enhanced or depleted stars, offering potential for advancing studies of stellar populations and Galactic archaeology.

Beyond its application to synthetic low-resolution spectra as a proof of concept, our approach is highly general and can be applied to any large-scale spectroscopic survey. In particular, higher-resolution datasets—such as those from APOGEE—are likely to benefit even more from this architecture. Increased spectral resolution provides more detailed information about individual absorption features, which should enhance both the reconstruction accuracy and the specialization of the latent space to elemental abundances. The main required adaptation is in the choice of spectral regions assigned to each decoder, which can consist of individual absorption lines or broader spectral intervals, depending on the resolution and the chemical feature of interest. 
This flexibility makes our method a promising tool for analyzing stellar populations across a wide range of observational contexts.

While our synthetic spectra provide a controlled data set for training and evaluation, they are subject to certain limitations when transfering the method to real data. First, the atmospheric models used do not account for carbon or $\alpha$-elements in the model structure, meaning that the spectra may be less sensitive to variations in these abundances compared to real observations. 
Some of the cooler models in our data set may reach temperatures where the completeness and accuracy of the VALD linelist begin to break down, due to the prominence of molecular bands that can result in missing or inaccurate features \citep{Piskunov1995vald, Ryabchikova2015vald}.

Additionally, the synthesis may not fully capture the complexities of metal-poor stars, where departures from local thermodynamic equilibrium (LTE) and 3D effects become significant  \citep[e.g.,][]{Asplund99confrontation, Asplund01effects,Jofre2014gaia}. 
As a result, the synthetic spectra are likely less sensitive to chemical variations than real data, especially for carbon-enhanced and metal-poor stars (see, e.g. \citealt{Eitner2025m3dis}).

Furthermore, in real observations, certain spectral bins may be flagged as unreliable, for example, due to the presence of cosmic rays or other detector artifacts, and these regions are typically masked during analysis. Since our current model requires complete spectral inputs for effective processing, missing or unreliable data must be addressed. One approach is to impute the missing values, essentially filling in the flagged regions with estimates based on surrounding data. Alternatively, the model itself could be adapted to handle missing data directly, allowing it to learn how to ignore or properly interpret these unreliable regions without requiring explicit imputation (\emph{e.\,g.} \citealt{dupuy2024mdae}).
This flexibility would ensure that the model can work with real-world data while maintaining the accuracy of its learned representations.
Moreover, in VAE- or AE-based approaches, artifacts and other non-systematic effects are naturally filtered out during training (provided the training data set is sufficiently large and diverse) since such effects do not contribute to the systematic spectral variation \citep[\emph{e.\,g.},][]{Aquilue2025eeg}. As a result, when such artifacts are encountered during inference, they typically produce large reconstruction errors, as we observed in Sect.~\ref{sec:snr}. 
Another important direction is the treatment of uncertainties in both spectra and conditioning parameters. Our reconstruction loss already incorporates measurement errors through a weighted log-likelihood (Eq.~\ref{eq:rec}), which down-weights noisy fluxes and ensures that the training focuses on well-measured regions. A similar strategy could be extended to the auxiliary parameters \tefflogg in partially supervised settings, by incorporating their uncertainties as weights in the corresponding loss terms. Also, while in this work the auxiliary latent space $\vec{w}$ is learned deterministically, one could instead model it probabilistically by predicting both mean and variance and sampling from this distribution, in analogy to the chemical latent space $\vec{z}$, which is already treated variationally in our model.

In future work, we foresee several avenues for applying, extending, and refining this research: 
\begin{itemize} 
    \item Observational application and high-resolution follow-up: 
    As a simple proof of concept, we have tested the framework on LAMOST low-resolution data (Appendix~\ref{app:lamost}), showing that it can be trained on real survey spectra. However, broader application requires additional caution, particularly regarding data selection, artifact removal, and label reliability.
    The natural next step is to extend this to larger and more diverse spectroscopic surveys, enabling the identification of candidate chemically enhanced or depleted stars. 
    These candidates could then be targeted for high-resolution spectroscopic follow-up, allowing for the precise determination of chemical abundances and a deeper understanding of the accretion events and enrichment processes that have shaped the Milky Way.
    \item Improved chemical modeling: The model can be readily extended to include additional chemical elements. This can be achieved in any large scale survey, of any resolution. We expect that the model will be even more effective in high-resolution data sets, where the spectral features are more pronounced and the disentangling and alignment of different chemical components is clearer.
    \item Cross-domain applications: The core principles behind our approach—structured latent spaces and disentangled representations—are not specific to stellar spectroscopy. 
    Similar architectures could be adapted to other domains where data reflect multiple, interacting sources of variation.
    Examples might include atmospheric retrievals in exoplanet spectroscopy, modeling galaxy spectral energy distributions, or identifying drivers of variability in time-domain surveys of variable stars and transients.
\end{itemize}
Overall, our approach demonstrates that a physics-constrained, data-driven framework can effectively disentangle chemical information from stellar spectra, producing interpretable representations that align with physical drivers of variation. 
Compared to existing pipelines that map spectra directly to pre-computed labels, our framework offers a complementary, interpretability-driven perspective.
Rather than fitting the data to synthetic models or externally calibrated labels, it learns how spectra encode chemical information in a self-consistent, low-dimensional space guided by physical constraints to compensate for the lack of complete supervision.
The result is a step towards model-free abundance estimation, where chemical information emerges naturally from the data themselves: each latent feature corresponds to the direction that best reconstructs the observed spectra, rather than the one that best matches a theoretical model.
This makes the framework particularly promising for Galactic Archaeology studies, facilitating the exploration of regions of chemical space where models or calibrations are uncertain (such as metal-poor or chemically peculiar stars) and the discovery of new chemical structures and rare stellar populations directly from the data, without relying on existing grids or supervised training labels, where such peculiar stars might not be included.
Applying the framework reliably to real data, however, will require additional validation and tuning, for example to handle survey-specific systematics and calibrations, which are beyond the present scope.

It is also important to note that in the total absence of chemical labels, interpretation of the learned features is not trivial: interpreting necessarily requires some knowledge; otherwise, one is merely looking. In practice, a minimal calibration (such as a simple linear regression step to rescale latent features into physical abundance units or to determine the direction of correlation) is sufficient. This step does not impose a model but instead provides a physical reference that makes the learned manifold interpretable and comparable across surveys.

As spectroscopic surveys continue to grow in scale and quality, we believe that such methods can offer a powerful alternative philosophy to fully supervised pipelines for extracting chemical information and advancing our understanding of stellar and Galactic evolution. 
We will thus make our code publicly available, providing a foundation for future research in unsupervised stellar spectroscopy.

\section{Data availability}
All code accompanying this manuscript will be made available upon publication on GitHub at \url{https://github.com/theosig/stellar-physRL}.
\begin{acknowledgements}
   TS acknowledges financial support from Inria Chile ANID project CTI230007. PJ acknowledges partial financial support of FONDECYT Regular Grant Number 1231057. The authors thank Bruno Cavieres for his contributions during the initial stages of this project, Pablo Aníbal Peña Rojas for his helpful assistance, and the referee for a very constructive review of the paper.
\end{acknowledgements}
\bibliographystyle{aa}
\bibliography{references}
\begin{appendix}
\section{Losses and metrics}\label{app:losses_metrics}
This appendix describes the loss functions and metrics used throughout training and evaluation. 
Let $x_{ij}$ be the value of feature $j$ of data sample $i$, and let $\tilde{x}_{ij}$ be the corresponding reconstructed feature. Each data sample consists of $d$ features, and $n$ is the number of data samples under consideration. 
The uncertainty on $x_{ij}$ is denoted by $\sigma_{ij}$. 
The latent space has dimension $d'$ (in this work $d'=m=3$), with $\mu_{ik}$ and $\epsilon_{ik}$ denoting the mean and standard deviation of the $k^{\text{th}}$ latent feature of data sample $i$.

$\mathcal{L}_{\rm rec}$ represents the reconstruction loss, which we choose to implement as the log-likelihood loss:
\begin{align}
    \mathcal{L}_{\rm rec} = -\frac{1}{2n} \sum_{i=1}^{n}{\left[ \log(2\pi) + \frac{1}{d} \sum_{j=1}^d \left(\log \sigma_{ij}^2 + \frac{(\tilde{x}_{ij} - x_{ij})^2}{\sigma_{ij}^2}\right) \right]},
    \label{eq:rec}
\end{align}
This loss gives more weight to precise measurements and less to noisy ones.

$\mathcal{L}_{\rm KL}$ is the Kullback–Leibler divergence between the approximate posterior 
$q(\vec{z}|\vec{x})=\mathcal{N}(\mu, \Sigma_q)$ and the prior distribution 
$p(\vec{z}) = \mathcal{N}(0, \Sigma_{\rm prior})$ \citep{Csiszar1975divergence}:
\begin{equation}
\mathcal{L}_{\rm KL} 
= \frac{1}{2n} \sum_{i=1}^{n} \Big[ 
\log \frac{\det \Sigma_{\rm prior}}{\det \Sigma_{q,i}}
- d' + \mathrm{Tr}\!\big(\Sigma_{\rm prior}^{-1}\Sigma_{q,i}\big) 
+ \mu_i^\top \Sigma_{\rm prior}^{-1} \mu_i \Big],
\label{eq:KL}
\end{equation}
where $\mu_i$ is the mean vector for sample $i$, 
and $\Sigma_{q,i} = L_i L_i^\top$ is its covariance matrix obtained 
via the Cholesky factor $L_i$ produced by the encoder.

In this work, $\Sigma_{\rm prior}$ is not the identity matrix but a fixed correlation matrix with unit diagonal elements and non-zero off-diagonal terms, encoding prior knowledge about correlations between abundances. 
For example, we set correlations $\rho_{\mathrm{Fe},\alpha}=-0.3$ and $\rho_{\mathrm{Fe},C}=-0.17$, while keeping other entries zero. 
This design allows the latent space to reflect correlations between elements, while preserving unit variance for each dimension.
For reference, the common special case of a diagonal posterior $q(\vec{z}|\vec{x})=\mathcal{N}(\mu, \mathrm{diag}(\epsilon^2))$ and an isotropic prior $p(\vec{z})=\mathcal{N}(0,I)$ reduces Eq.~\ref{eq:KL} to:
\begin{equation}
\mathcal{L}_{\rm KL} = \frac{1}{2n} \sum_{i=1}^{n} \sum_{k=1}^{d'} \left(\mu_{ik}^2 + \epsilon_{ik}^2 - \log \epsilon_{ik}^2 - 1 \right),
\label{eq:KL_diag}
\end{equation}
which is the standard expression often used in VAEs.

$\mathcal{L}_{\rm class}$ is the classification loss defined as the sparse categorical cross entropy
\begin{equation}
    \mathcal{L}_{\text{class}} = -\sum_{i=1}^n\sum_{l=1}^q y_{il}\log(p_{il})\,,
\end{equation}
where $y_{il}$ is the true label of the sample $i$ and $p_{il}$ is the predicted probability for it belonging to the $l^{th}$ class.
The encoder-decoders are trained to maximize this $\mathcal{L}_{\text{class}}$ (in order to have a representation which is uninformative about the non-chemical physical parameters), as shown by the minus sign in Eq.~\ref{eq:objective}, while the classifier is trained in order to minimize it.
After training we evaluate the reconstruction performance using the relative $L^2$ error:
\begin{equation}
    \text{Relative } L^2 \text{ error} = \frac{\sum_{i=1}^n \sum_{j=1}^d (\tilde{x}_{ij} - x_{ij})^2}{\sum_{i=1}^n \sum_{j=1}^d x_{ij}^2}.
    \label{eq:rel_l2}
\end{equation}

Two metrics are used to evaluate the classification performance for identifying chemically enhanced or depleted stars. 
Let $c$ denote the positive class, i.e., the chemically enhanced or depleted stars of interest:
\begin{enumerate}
    \item The $\text{recall}(c)$ represents the fraction of true positive-class stars that are correctly identified: 
    \begin{align}
        \text{recall}(c)=
        \frac{\text{True Positives}(c)}{\text{True Positives}(c) + \text{False Negatives}(c)}.
        \label{eq:recall}
    \end{align}
    A high recall indicates that most positive-class stars are successfully detected, whereas a low recall suggests many are missed.
    \item The $\text{precision}(c)$ indicates the fraction of stars predicted to belong to class $c$ that truly do:
    \begin{align}
        \text{precision}(c)=
        \frac{\text{True Positives}(c)}{\text{True Positives}(c) + \text{False Positives}(c)}.
        \label{eq:precision}
    \end{align}
    High precision means that most stars flagged as belonging to class $c$ are indeed correct. Precision is sometimes referred to as purity.
\end{enumerate}

\section{Hyperparameters}\label{app:training_details}
In Table \ref{table:hyperparameters} we detail the hyperparameters used in our training setup.
We note that the specific architecture parameters (such as the number and size of encoder layers) had a relatively minor effect on the overall performance in our experiments.
The batch size was chosen to ensure that each batch fits into GPU memory during training, but it can be increased on systems with more available memory.

\begin{table}
\centering       
\caption{Hyperparameters configuration.}
\begin{tabular}{p{0.48\linewidth}p{0.48\linewidth}}
\toprule         
\textbf{Parameter} & \textbf{Value}\\
\midrule
\textbf{Architecture Parameters}&\\
Encoder convolutional layers size & (128, 256, 512) \\
Encoder convolutional filters size & (5, 11, 21) \\
Encoder fully connected layers size & (128, 64, 32) \\
Decoder layers size & (32, 64, 128) \\
\textbf{Training Parameters}&\\
Optimizer & Adam \citep{kingma2014adam}\\
Maximum learning rate & $10^{-3}$\\
Learning rate scheduler& 1Cycle \citep{smith2019super}\\
Training steps & $4000$\\
Batch size & $512$ \\
&\\
\textbf{Weighting Parameters} (Eq.~\ref{eq:objective})&\\
 
$\beta$ & $5 \times 10^{-3}$ \\
$\lambda$ & $\begin{cases}
0, & \text{if } \text{step} \leq 200\\
10^{-2}, & \text{if } 200 < \text{step} \leq 400\\
10^{-1}, & \text{if } 400 < \text{step} \leq 1000\\
1, & \text{if } \text{step} > 1000
\end{cases}$ \\
\bottomrule
\end{tabular}
\tablefoot{Step refers to the number of training epochs (full passes through the dataset), not individual gradient updates. The annealing schedule for $\lambda$ is designed to allow the network to focus on reconstruction early on, before gradually introducing the disentanglement term.}
\label{table:hyperparameters}
\end{table}
\section{Effect of disabling selective gradient flow}\label{app:gradflow}
To assess the impact of selective gradient flow (Sect.~\ref{sec:gradients}), we trained an otherwise identical model without restricting decoder gradients. In this setup, each decoder can backpropagate its reconstruction loss through all latent dimensions. 

\begin{figure}
    \centering
    \includegraphics[width=\linewidth]{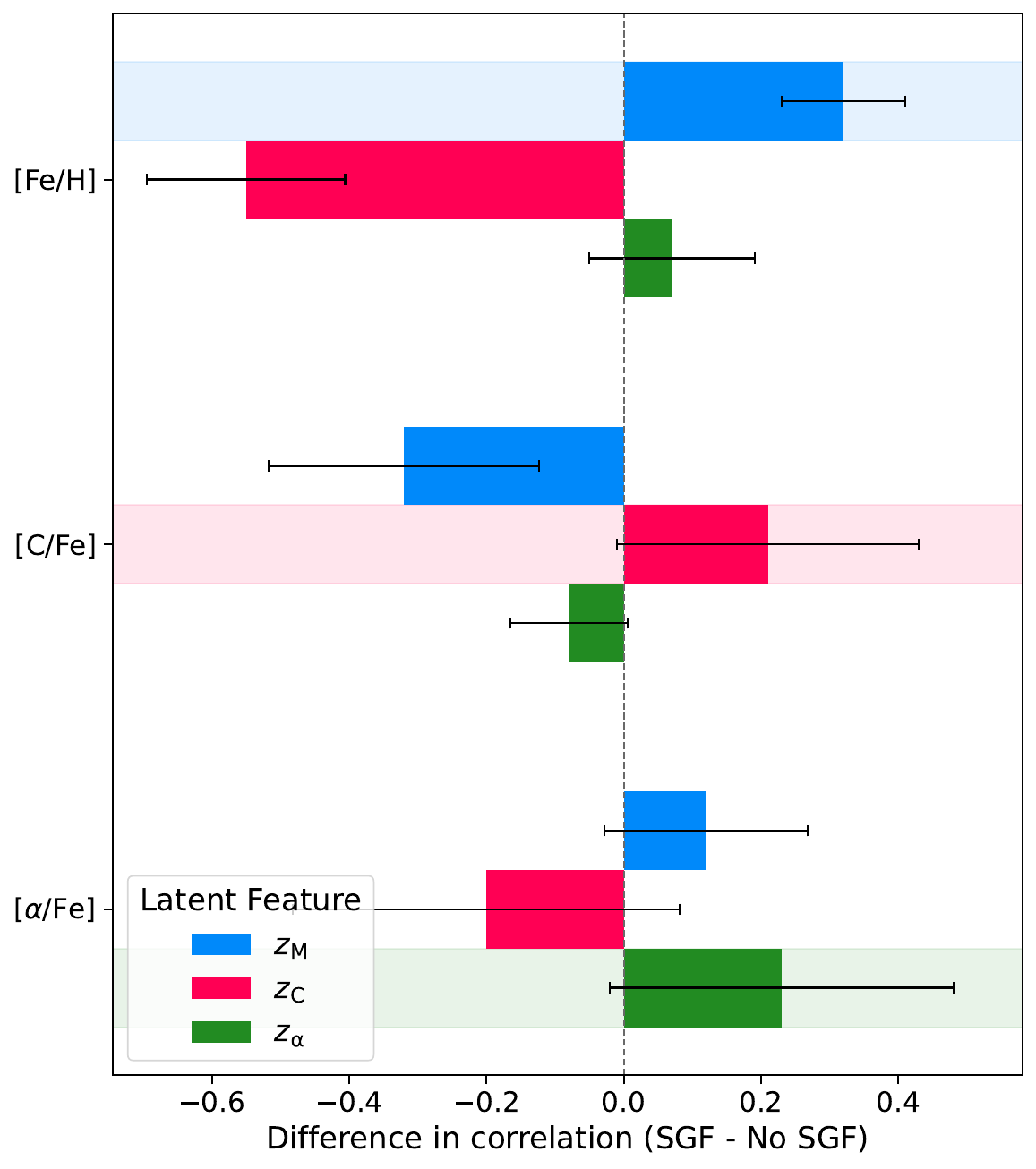}
    \caption{Gain in correlations when the model is trained with and without selective gradient flow (SGF and no SGF, respectively), color–coded according to the legend. The shaded areas highlight the latent feature associated with each chemical label. Positive values for the corresponding feature, combined with negative values for the others, indicate that selective gradient flow has a strong positive impact.}
    \label{fig:gain_corr}
\end{figure}

Table~\ref{table:gradflow_corr} reports the average absolute Pearson correlation coefficients between each latent feature and the stellar parameters, computed over multiple random realizations of training and validation sets when selective gradient flow is not applied. The corresponding differences in correlations obtained with and without selective gradient flow are shown in Fig.~\ref{fig:gain_corr}.

We observe that: (i) the mapping between each latent and its nominal chemical abundance (e.g. $z_{\rm M}$ with [Fe/H]) becomes weaker; (ii) off-diagonal correlations (e.g. $z_{\rm C}$ vs [Fe/H], $z_{\rm M}$ vs [C/Fe]) increase, indicating stronger mixing of chemical information across latent dimensions; and (iii) the dispersion of these correlations across random realizations is larger, implying that the latent to abundance mapping is less stable from run to run.

The average relative $L^2$ error on the validation set is slightly lower in this case ($0.004$ vs.\ $0.006$ with selective gradient flow). This can be explained by the additional freedom gained when gradient constraints are removed, allowing the model to further minimise the reconstruction loss.

We conclude that disabling selective gradient flow improves raw reconstruction but undermines both the interpretability and the robustness of the latent axes: they become more entangled and less repeatable across random seeds. As expected, selective gradient control therefore promotes a clearer and more stable mapping between latent features and their targeted chemical abundances, at the cost of a negligible penalty in reconstruction quality.

\begin{table}
\centering    
\caption{Correlations of the latent features with atmospheric parameters, without selective gradient flow.}
\begin{tabular}{lccc}          
\toprule          
Target feature & \textbf{$z_{\rm M}$} & \textbf{$z_{\rm C}$} & \textbf{$z_{\rm \alpha}$}\\
\midrule
$\rm [Fe/H]$ 
  & \makecell{$0.60\pm0.09$* \\ (\corrFE{})} 
  & \makecell{$0.60\pm0.12$ \\ ($0.05\pm0.08$)} 
  & \makecell{$0.23\pm0.05$ \\ ($0.30\pm0.11$)} \\
  
$\rm [C/Fe]$ 
  & \makecell{$0.65\pm0.18$ \\ ($0.33\pm0.08$)} 
  & \makecell{$0.71\pm0.22$* \\ (\corrC{})} 
  & \makecell{$0.17\pm0.08$ \\ ($0.09\pm0.03$)} \\
  
$\rm [\alpha/Fe]$ 
  & \makecell{$0.21\pm0.13$ \\ ($0.33\pm0.07$)} 
  & \makecell{$0.29\pm0.28$\\ ($0.09\pm0.03$)} 
  & \makecell{$0.59\pm0.25$* \\ (\corra{})} \\
  
$T_{\rm eff}$ 
  & \makecell{$0.05\pm0.02$ \\ ($0.03\pm0.01$)} 
  & \makecell{$0.04\pm0.02$ \\ ($0.01\pm0.01$)} 
  & \makecell{$0.03\pm0.01$ \\ ($0.03\pm0.02$)} \\
  
$\log{g}$ 
  & \makecell{$0.07\pm0.06$ \\ ($0.03\pm0.01$)} 
  & \makecell{$0.07\pm0.06$ \\ ($0.01\pm0.01$)} 
  & \makecell{$0.02\pm0.01$ \\ ($0.03\pm0.02$)} \\
\bottomrule
\end{tabular}
\tablefoot{Average absolute Pearson correlation coefficients $r$ between each latent feature and stellar parameters, when the model is trained without selective gradient flow. Values in parentheses indicate the reference results obtained with selective gradient flow (Table~\ref{table:correlations}). * highlights the correlation between each chemical abundance and its associated latent feature.}
\label{table:gradflow_corr}
\end{table}

\section{Spectral regions}\label{app:regions}
In this appendix, we report the spectral regions associated with each decoder, selected based on the presence of strong lines or bands characteristic of each corresponding element. Table \ref{table:alpha_regions} presents the regions for the $\alpha$-elements decoder, while Table \ref{table:carbon_regions} shows the regions for the carbon decoder.
\begin{table}[ht]
    \centering       
    \caption{Target spectral regions for the $\alpha$-elements decoder.}
    \begin{tabular}{lc}
        \toprule
        Feature & Region (Å) \\
        \midrule
        Ca I & $4215-4240$\\
        Mg I & $4568-4574$\\
        Mg Region & $5035-5051$, $5160-5190$ \\
        Si I & $6150-6170$\\
        O I & $6290-6310$\\
        TiO Bands & $7142-7154$\\
        Ca II Infrared Triplet & $8493-8503$, $8534-8550$,\\
        & $8655-8669$ \\
        Ti I Lines & $8405-8445$ \\
        S I & $9205-9240$\\
        \bottomrule
    \end{tabular}
    \label{table:alpha_regions}
\end{table}

\begin{table}[ht]
    \centering       
    \caption{Target spectral regions for the carbon decoder.}
    \begin{tabular}{lc}
        \toprule
        Feature & Region (Å) \\
        \midrule
        CH G Band & $4280-4320$\\
        C2 swan 4737 & $4730-4745$\\
        C I & $5050-5054$ \\
        C2 swan 5165 & $5150-5180$\\
        C2 swan 5635 &$5625-5645$\\
        C I 6588& $6585-6590$\\
        C I 8727 & $8725-8729$\\
        \bottomrule
    \end{tabular}
    \label{table:carbon_regions}
    \end{table}

\section{Abundance estimation from learned representations}\label{app:inference}
While not intended as a replacement for supervised models (at least not yet), we assess the capabilities of our representation to perform direct abundance estimation. To do so, and to provide context for its performance, we compare it with The Cannon \citep{ness2015cannon}, a widely used supervised method for abundance analysis (\emph{e.\,g.}, \citealt{wheeler2020abundances,walsen2024assembling,manea2024chemical}). 
We adopt this method because of its simplicity and fast training, which makes it well suited for a quick sanity check. More complex deep-learning methods (e.g. astroNN) require substantial training and tuning, which are outside the scope of this work.

We trained The Cannon on $1\,000$ randomly selected spectra (with the same SNR as used in our VAE experiments) using exact, unperturbed labels, and evaluated it on a test set of $1\,000$ random spectra. The resulting root mean squared errors (RMSEs) for $\rm [Fe/H]$, $\rm [C/Fe]$, and $\rm [\alpha/Fe]$ are $0.26$, $0.14$, and $0.35$ dex, respectively.
Our representation learning model, by contrast, is trained in a self-supervised fashion with no access to chemical abundance labels. It is, however, conditioned on stellar parameters ($T_{\rm eff}$ and $\log g$), which are treated as known inputs during training. Using then a simple supervised univariate linear regression from the learned latent features to chemical abundances, we obtained lower RMSEs of $0.21$ dex for $\rm [Fe/H]$, $0.08$ dex for $\rm [\alpha/Fe]$, and $0.31$ dex for $\rm [C/Fe]$. 

In fairness to The Cannon, we note that it must simultaneously learn all stellar labels—including $T_{\rm eff}$ and $\log g$—while our method disentangles chemical information directly from spectra under the assumption that these physical parameters are already known. In our approach, we first learn a latent representation conditioned on $T_{\rm eff}$ and $\log g$, and abundance prediction is performed only afterward via a separate regression step. Moreover, we used The Cannon with default settings and a randomly selected training set; with more careful tuning or data selection, its performance is likely to improve.

Regardless, despite the simplicity of the regression step and the absence of chemical labels during representation learning, our model achieves competitive performance in abundance estimation—even with a simple univariate linear regressor that can, in principle, be trained on as few as two labeled examples per abundance.
The regression step primarily serves to rescale each latent feature into dex, effectively acting as a lightweight calibration.
Although this is not a direct comparison—due to differing assumptions and training setups—these results highlight the potential of our method to support chemical abundance estimation in regimes where supervision is sparse.

\section{Effect of SNR and training set size on reconstruction error}\label{app:noise}
Variational autoencoders are known to act as denoising models: the low-dimensional latent bottleneck prevents the network from memorizing high-dimensional random noise and forces reconstructions to lie close to the underlying data manifold. This implies that, when trained on a large enough number of noisy spectra, the model should still recover cleaner representations than the inputs themselves.

\begin{figure}
    \centering
    \includegraphics[width=0.48\textwidth]{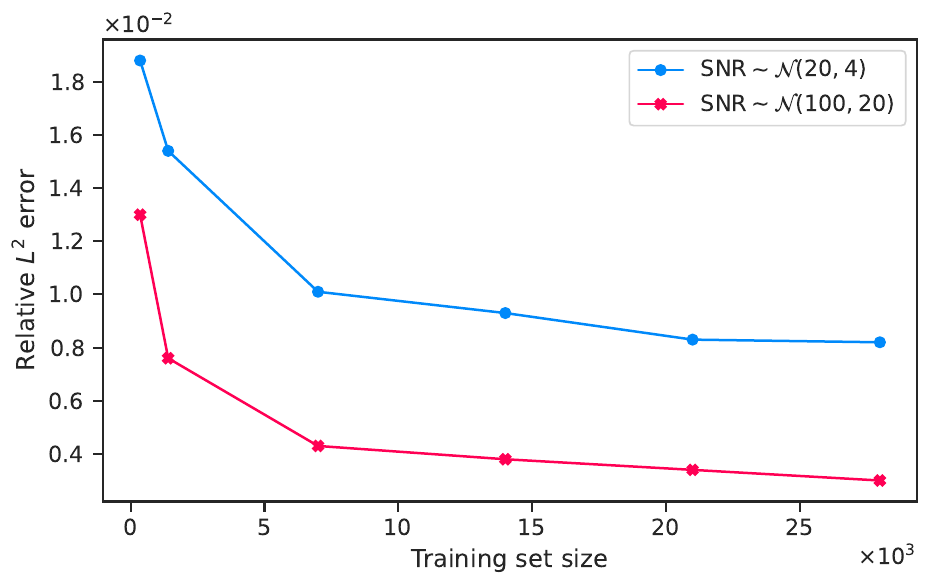}
    \caption{Reconstruction error as a function of training set size for spectra with SNR $\sim 20$ (blue) and $\sim 100$ (pink).}
    \label{fig:snr_scaling}
\end{figure}

To verify this property in our setting, we trained the model separately on spectra with low SNR ($\sim \mathcal{N}(20,4)$) and high SNR ($\sim \mathcal{N}(100,20)$), and evaluated the reconstruction error with respect to the corresponding clean synthetic spectra (before noise was added) as a function of training set size. The results are shown in Fig.~\ref{fig:snr_scaling}: blue for training with low-SNR data, and pink for high-SNR. In both cases, the reconstruction error remains below the input noise variance, confirming that the model does not reproduce random fluctuations. The denoising effect becomes stronger with larger training sets, as the manifold is more densely sampled, but is already visible with small data sets. We also note that the reconstruction error for models trained on low-SNR spectra never reaches the level achieved with high-SNR training, indicating that while the model learns cleaner representations than the noisy inputs, it does not completely ignore the noise. 
Moreover, this experiment assumes purely Gaussian noise; real observations may contain additional systematics or non-Gaussian effects that could further affect reconstruction quality. In this context, advanced strategies such as data augmentation or explicit noise modeling could help the model better disentangle signal from noise and reduce the performance gap between low- and high-SNR regimes, although exploring these directions is beyond the scope of the present work.

\section{An experiment on LAMOST data}\label{app:lamost}
\begin{figure}
    \centering
    \includegraphics[width=\linewidth]{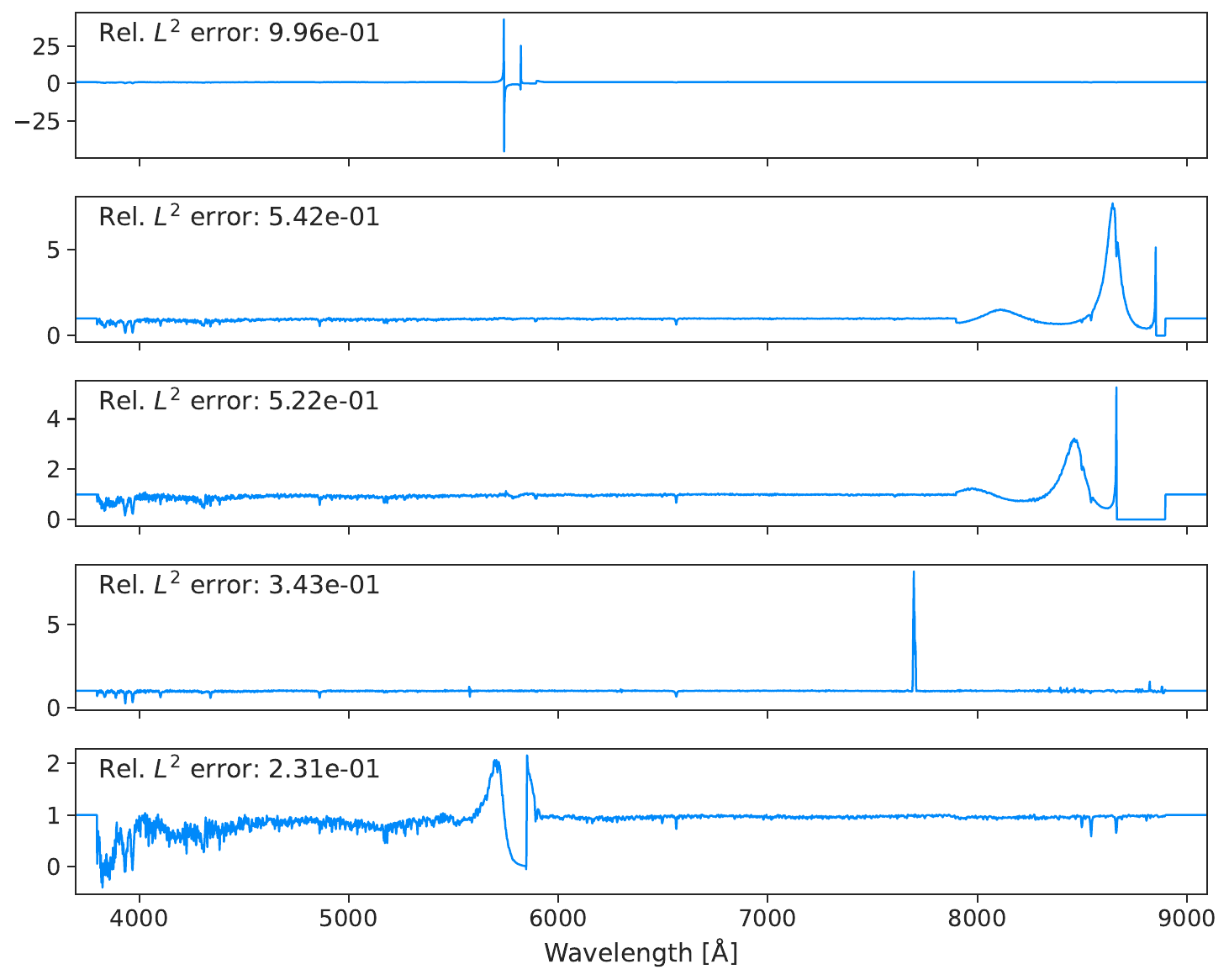}
    \caption{LAMOST DR10 normalized spectra with highest reconstruction errors.}
    \label{fig:lamost_bad}
\end{figure}
\begin{figure*}
    \centering
    \includegraphics[width=\linewidth]{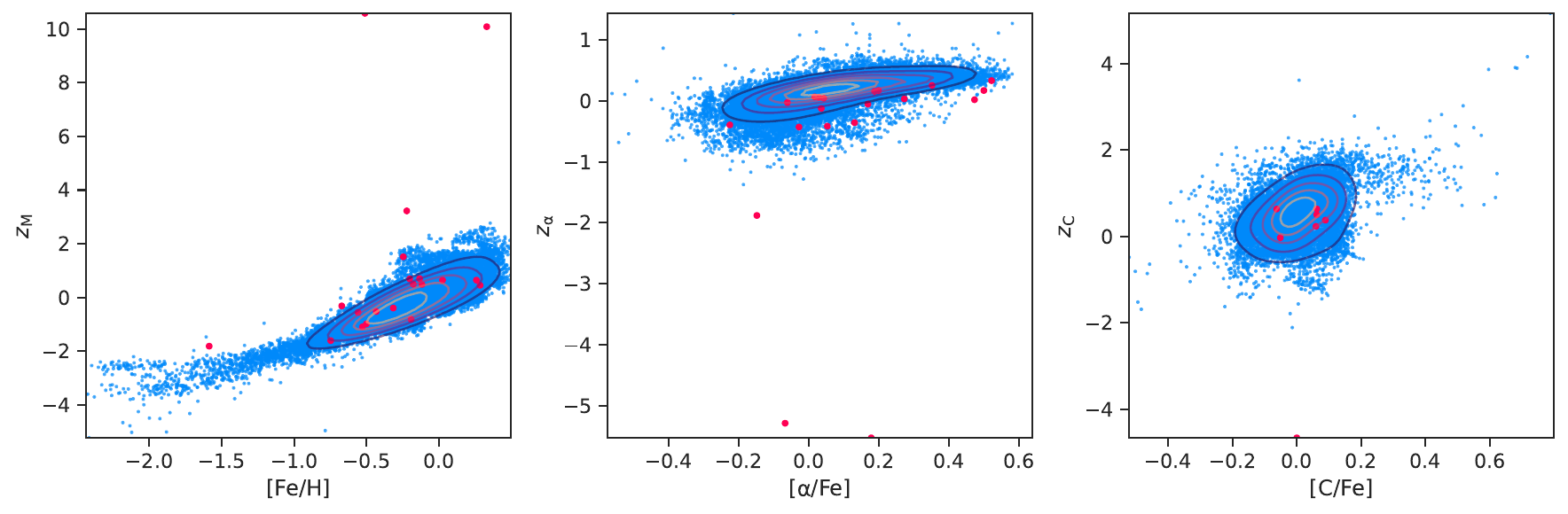}
    \caption{
    Contour plot of latent features (from left to right, $z_{\mathrm M}, z_{\mathrm \alpha}, z_{\mathrm C}$) and their corresponding chemical abundances for LAMOST spectra. The scatter points represent individual data points, and the contour lines represent data density, with lighter contours indicating regions of higher density. In red we show the 20 spectra with largest reconstruction errors.}
    \label{fig:datalatent}
\end{figure*}
To test the applicability of our method to real survey spectra, we performed an experiment on LAMOST DR10 low-resolution (LRS) data \citep{cui2012lamost}. 
We selected stars with no spectral flags, and required that the first 250 flux points (corresponding to the bluer part of the spectrum) have ${\rm SNR}>20$.
We adopted stellar parameters $T_{\rm eff}$, $\log g$, ${\rm [Fe/H]}$, and ${\rm [\alpha/Fe]}$ from the LAMOST stellar parameter pipeline (LASP, \citealt{Wu2014automatic}), restricting the sample to $T_{\rm eff}\in[3000,7000]$ K, $\log g\in[2.5,6]$ dex, and relative errors on $T_{\rm eff}$ below $3\%$.

After correcting the spectra for radial velocity, we interpolated them onto a common grid between 4000 and 9000\,\AA.  This selection resulted in a sample of $\sim$40,000 stars. For carbon abundances (not provided by LASP), we adopted convolutional neural networks-derived labels from \citet{liu2025cnn}. 
We used the same model and wavelength regions as in the synthetic experiments (see Sect. \ref{app:regions}), excluding regions outside the LAMOST coverage.
Training the model from scratch on this data set, we evaluated the same diagnostics as in the synthetic case:  
\begin{itemize}
    \item The average relative $L^2$ error is $0.013$, very close to the value obtained for synthetic noisy spectra ($0.014$; in the main body of the paper we reported the error with respect to the original, unperturbed spectra). Figure~\ref{fig:lamost_bad} shows the input spectra of the stars with the largest reconstruction errors. Many of these spectra display strong artifacts that were not flagged in the survey pipeline, suggesting that additional quality cuts or artifact removal would improve training. Importantly, the model also retains its capability to recognize anomalous spectra, as discussed in Sect.~\ref{sec:snr}, where high reconstruction errors naturally highlight outliers and artifacts.
    \item The auxiliary latent space $\vec{w}$ retains predictive power for $T_{\rm eff}$ and $\log g$, with RMSE of $\sim0.12$, compared to $0.11$ in the synthetic case. These values are computed on features rescaled to zero mean and unit variance, so they should be interpreted in relative rather than absolute physical units.
    \item The latent dimensions correlate with LAMOST metallicities, $\alpha$-element abundances, and the CNN-derived carbon abundance, though less strongly than in the synthetic experiments. Specifically, $z_{\rm M}$ correlates with [Fe/H] at $r \approx 0.89$, $z_{\alpha}$ with [$\alpha$/Fe] at $r \approx 0.65$, and $z_{\rm C}$ with [C/Fe] at $r \approx 0.40$. Figure~\ref{fig:datalatent} illustrates these relations, highlighting in red the 20 spectra with the largest reconstruction errors. 
    While the relation between $z_{\rm C}$ and [C/Fe] is noisy, stars with higher [C/Fe] values still occupy the high end of $z_{\rm C}$, indicating that the model might still be extracting some meaningful carbon information. We note, however, that individual abundance measurements in LAMOST—particularly those derived from external CNN-based estimates—can be subject to biases (e.g., \citealt{Ho2017label}) and substantial uncertainties, which may contribute to the weaker observed correlation. In contrast, [Fe/H] is typically the most reliable abundance measurement in low-resolution surveys, as it is constrained by numerous features and is the primary calibration target of most survey pipelines, which likely explains its stronger correlation with $z_{\rm M}$.
\end{itemize}
In this test, we found that stopping training at $\sim$1000 iterations gave the best results. While some additional hyperparameter tuning might further improve performance, this is outside the scope of the present work.

Overall, the model can be trained successfully on real low-resolution survey data, achieving reconstruction quality comparable to the synthetic case. The experiment also highlights the importance of careful data selection and validation: enforcing SNR cuts (especially in the blue), filtering out spectra with artifacts (as seen in the largest-error cases), and checking survey products such as continuum normalization and object type classifications. The weaker correlations between latent features and abundances likely reflect limitations in the survey-derived labels themselves, which vary in reliability across elements and pipelines. This does not diminish the utility of the synthetic experiments presented in the main text, which provide a controlled proof of concept. Rather, it shows that fully exploiting and validating such methods on real survey data requires additional care in sample selection, artifact handling, label reliability and hyperparameter tuning.
\end{appendix}

\end{document}